\documentclass[aps,showpacs,12pt,amssymb,preprintnumbers,amsmath,amsfonts]{revtex4}
\usepackage{latexsym}

\newcommand{\ba}{\begin{eqnarray}}
\newcommand{\ea}{\end{eqnarray}}

\newcommand{\be}{\begin{equation}}
\newcommand{\ee}{\end{equation}}
\newcommand{\pa}{\partial}

\newcommand{\nn}{\nonumber}

\newcommand{\Om}{\mathbf{\Omega_p}}

\newcommand{\Bvec}{\mathbf{B}}

\usepackage{dcolumn}
\usepackage[dvips]{graphicx}
\usepackage{color}
\usepackage{slashed}

\begin{document}
\title{Dynamical evolution of the chiral magnetic effect:\\ Applications to the quark-gluon plasma}
\author{Cristina Manuel}
\affiliation{Institut de Ci\`encies de l'Espai (IEEC/CSIC), Campus Universitat Aut\`onoma
de Barcelona, Facultat de Ci\`encies, Torre C5, E-08193 Bellaterra, Spain}
\author{Juan M. Torres-Rincon}
\affiliation{Subatech, UMR 6457, IN2P3/CNRS, Universit\'e de Nantes, \'Ecole de Mines de Nantes, 4 rue Alfred Kastler 44307,
Nantes, France}
\pacs{12.38.Mh,11.30.Rd}

\begin{abstract} 
 We study the dynamical evolution of the so-called chiral magnetic effect in an electromagnetic conductor.
To this end, we consider the coupled set of corresponding Maxwell  and chiral anomaly equations, and  we prove that
these can be derived from chiral kinetic theory.  After integrating the chiral anomaly equation over space in a closed
volume,  it  leads to a quantum conservation law of the total helicity  of the system. A change in the magnetic helicity density comes together
with a modification of the chiral fermion density. We study in Fourier space the  coupled set of anomalous equations and 
we obtain the dynamical evolution of the magnetic fields, magnetic helicity density, and chiral fermion imbalance. Depending on the initial conditions we observe how the
helicity might be transferred from the fermions to the magnetic fields, or vice versa, and find that the rate  of this transfer 
also depends on the scale of  wavelengths of the gauge fields in consideration.  We then focus our attention on the quark-gluon plasma phase
and analyze the dynamical evolution of the chiral magnetic effect in a very simple toy model.  We conclude that an existing chiral fermion
imbalance in peripheral heavy ion collisions would affect the magnetic field dynamics and, consequently, the charge-dependent correlations measured in these experiments.
\end{abstract}


\maketitle

\section{Introduction}

Quantum chiral anomalies have a serious impact on the macroscopic physics of finite systems made by (quasi) massless
fermions. One specific example of this fact is the so-called
chiral magnetic effect (CME), which refers to the generation of an electromagnetic (EM) current
proportional to the magnetic field in systems where there exists an imbalance between the populations
of right- and left-handed fermions~\cite{Kharzeev:2013ffa}. Initially discovered in electroweak
plasmas~\cite{Vilenkin:1980fu}, it was pointed out that the CME could be realized in noncentral heavy-ion collisions (HICs), where
large magnetic fields are generated, and the topological properties of QCD allow for the
generation of the chiral fermion imbalance ~\cite{Kharzeev:2007jp,Fukushima:2008xe}. The CME can also occur in condensed matter
systems, such as Weyl and Dirac semimetals with chiral quasiparticles~\cite{Son:2012bg,Basar:2013iaa}. Experimental observation
of the CME in a Dirac semimetal has been recently reported through measurements of its
magneto-transport properties~\cite{Li:2014bha}.

The CME has been reproduced with a variety of methods like quantum field theory~\cite{Kharzeev:2007jp,Fukushima:2008xe,Vilenkin:1980fu,Gorbar:2011ya}, holographic approaches~\cite{Yee:2009vw,Rebhan:2009vc,Gynther:2010ed} 
and lattice studies~\cite{Buividovich:2009wi,Abramczyk:2009gb}, where as an initial condition one assumes
a constant and homogeneous chiral fermionic imbalance in the presence of a magnetic  field.

A formulation of transport theory for chiral fermions ---that also describes effects of
the quantum chiral anomalies--- has been developed in Refs.~\cite{Son:2012wh,Stephanov:2012ki,Son:2012zy,Manuel:2013zaa,Manuel:2014dza}.  For example, this can be achieved
by considering the first $\hbar$-correction to the classical dynamics of chiral fermions. Quantum corrections to the
standard classical Boltzmann equation can be easily incorporated, and the resulting framework  not only describes 
correctly the chiral anomaly equation for finite temperature and
density fermionic systems, but also other anomalous effects, such as the CME. The basic requirement of the chiral kinetic theory (CKT) is the existence
of well-defined chiral quasiparticles.

 In this manuscript we study the dynamical evolution of the CME by deducing the form
of the electromagnetic (EM) current from the CKT within the relaxation time approximation, for times larger than
the collision time. Then, we study the Maxwell equations which have this EM current as a source, coupled to the
chiral anomaly equation, which is also deduced from CKT. Similar coupled anomalous Maxwell equations (AMEs) have been first considered for small
wavelengths in cosmological contexts; see e.g., Refs.~\cite{Giovannini:1997eg,Banerjee:2004df,Boyarsky:2011uy,Tashiro:2012mf}. Here, they are derived in a more general way.

The chiral anomaly equation expresses a quantum conservation law. After integrating it over space in a closed
volume,  it relates the helicity of fermions with the Abelian Chern--Simons number or magnetic helicity (as known in magnetohydrodynamics) in such a way that a linear combination of these two quantities is conserved.
In the main part of this manuscript  ---regardless the method
used to derive the AME--- we study the dynamical evolution of the magnetic fields and of the magnetic helicity,
constrained by this quantum conservation law imposed by the chiral anomaly. We assume that other processes that
might change  the fermion chirality are absent, or at least negligible in the range of time scales we consider. Depending on the initial conditions we observe how the 
helicity might be transferred from the fermions to the gauge fields, or vice versa, and find that the rate of this transfer 
also depends on the scale of wavelengths of the gauge fields in consideration.

While the above features are generic for any chiral fermionic electromagnetic conductor, we focus our interest in the
quark-gluon plasma (QGP) phase that is supposed to be realized in HICs. 
We employ a very simple toy model for the QGP, but that can be very illustrative to extract relevant conclusions. In particular, we assume a QGP in a
static finite volume, and consider the evolution of an existing chiral fermion imbalance in the presence of a large initial
magnetic field, with Gaussian shape, and no initial magnetic helicity.  We observe a rapid creation of magnetic helicity and a reduction of the fermion chiral
imbalance, of oscillatory behavior at very short time scales. We claim that the specific evolution of both chiral imbalance
and magnetic helicity depends on the initial spectrum of magnetic fields,  but the effects we found are quite generic and have been overlooked so far in studying the impact of the CME in HIC physics. Further, we
also consider the generation of a chiral fermion imbalance from magnetic field configurations with high magnetic helicity.

This manuscript is structured as follows.
The set of AMEs is studied in Sec.~\ref{sec:AME}. We rewrite the AMEs in Fourier space in Sec.~\ref{sec:FAME},
making use of circular polarization vectors, as then the magnetic helicity can be written in terms of the left- and right-handed polarization components
of the magnetic field. We show how the dynamical evolution of  each field component is different in the presence of a fermion chiral  imbalance.
Exact solutions for small wave numbers are found and displayed in Sec.~\ref{sec:smallk}, where we also show plots of the dynamical
evolution of the  axial chemical potential and the magnetic helicity.
In Sec.~\ref{sec:largek} we treat the case of large wave numbers, where explicit solutions of the AMEs can only be numerically obtained.
We point out that solutions of these equations show a very different behavior for large or small wavelengths. In Sec.~\ref{Sec-QGP} we focus on the QGP, and argue that even
when the non-Abelian gauge field contribution to the chiral anomaly equation should be considered, at late times the color dynamics can be safely ignored.
We present numerical results for the evolution of the chiral fermion imbalance and magnetic helicity assuming an initial Gaussian spectrum for the magnetic field.
We present our conclusions in Sec.~\ref{conclu}. 
In Appendix~\ref{app:secCKT} we give a quick review of the CKT and from it, derive the form of the EM current for a finite temperature plasma with an initial fermion
chiral imbalance, using the relaxation time approximation. In Appendix~\ref{app:numerics} we give details on how we have numerically solved the AME.
Finally, in Appendix~\ref{app:fliprate} we compute the helicity-flipping rate of Compton scattering and argue that this
process, together with the helicity change by Landau damping can be ignored in the context of HICs.
We use natural units ($\hbar = c = k_B = 1$), throughout the manuscript.

\section{Anomalous Maxwell equations}
\label{sec:AME}

In this section we focus on the set of dynamical equations that are derived from the chiral transport
approach. In essence, we study Maxwell equations with the EM current associated
to the chiral fermions, together with the chiral anomaly equation, which is
a direct consequence of the chiral transport equation.

 Similar equations have been used for electroweak plasmas in cosmological contexts in
Refs.~\cite{Giovannini:1997eg,Banerjee:2004df,Boyarsky:2011uy} by assuming the same form for the EM current as the one obtained here.
 For an electromagnetic conductor, it is possible to obtain a dynamical equation for the magnetic field. To this end we consider the Maxwell equation
\be \label{eq:Maxwell}
\frac{\partial {\bf E}}{\partial t} + {\bf J} = \nabla \times {\bf B} \ ,
\ee
where the EM current contains both Ohm's law and the CME as derived in Appendix~\ref{app:secCKT},
\be
\label{elcurrentsta2}
 {\bf J} =  \sigma  {\bf E} + \sum_{s= 1}^{N_s} \frac{e_s^2 \, \mu_5}{4 \pi^2} \, {\bf B} \ .
  \ee
We take the curl of Eq.~(\ref{eq:Maxwell}), and then use Faraday's law and the explicit form of the current in Eq.~(\ref{elcurrentsta2}) to reach to
\be
\label{eq-Bfield}
\frac{\partial {\bf B}}{\partial t} = \frac{1}{\sigma} \nabla^2 {\bf B} + \frac{C \alpha \mu_5}{\pi \sigma}  \nabla \times {\bf B} - \frac{1}{\sigma} \frac{\partial^2 {\bf B}}{\partial t^2} \ ,
\ee
where we have defined the factor
\be
C \equiv \sum_{s=1}^{N_s} \frac{e_s^2}{e^2} \ , 
\ee
 that takes into the account the sum of the charges squared of every fermion species in the plasma and $\alpha=e^2/(4 \pi)$ is the EM fine structure constant.

To arrive to Eq.~(\ref{eq-Bfield}), we have assumed that both the axial chemical potential $\mu_5$ and
the electrical conductivity $\sigma$ are homogeneous, or at least, that
their spatial variations can be neglected in front of the spatial variation of the magnetic field. Otherwise, gradients of these two quantities would correct Eq.~(\ref{eq-Bfield}).

The magnetic field evolution is then linked to the value of the chemical potential of the chiral imbalance. Thus Eq.~(\ref{eq-Bfield}) should be studied coupled to the
chiral anomaly equation, Eq.~(\ref{chiral-anomaly}). In a system with $N_s$ different fermion species, we integrate the total axial current over the volume $V$ to get
\be
\label{basicAnoeq}
\frac{d(n_R - n_L)}{dt} = \frac{2 C \alpha}{\pi} \frac{1}{V} \int d^3 x \, {\bf E}\cdot {\bf B} =  - \frac{ C \alpha}{\pi} \frac{d{\cal H}}{dt} \ ,
\ee
where ${\cal H}$ is the magnetic helicity density, defined as
\be
{\cal H} (t) =  \frac{1}{V} \int_V d^3 x \, {\bf A}\cdot {\bf B} \ .
\ee
This quantity is gauge invariant provided that ${\bf B}$  vanishes at, or is parallel to the boundary of $V$. As we will focus our study on closed systems, 
the use of the magnetic helicity as a dynamical variable will not bring any gauge dependence in our results. 
If we further take into account that there might be other processes that change the fermion chirality, the dynamical evolution of the chiral imbalance is then
governed by the equation
\be
\label{anomal-full}
\frac{d n_5}{dt} =  - \frac{ C \alpha}{\pi} \frac{d{\cal H}}{dt} - \Gamma_f n_5 \ ,
\ee
 where the helicity-flipping rate $\Gamma_f$ depends on the system under consideration. This coefficient will be extensively discussed later for a QGP.

It is also convenient to express the above equation in terms of the axial chemical potential. At very high temperature $T\gg \mu_5$, one writes 
\be
\label{eq-mu5}
\frac{d\mu_5}{dt} =  - c_\Delta C \alpha \frac{d{\cal H}}{dt}  - \Gamma_f \mu_5\ , \ee
where $c_\Delta = \frac{1}{ \pi  \chi_5}$  and $\chi_5 = \frac{\partial n_5}{\partial \mu_5}$ is the susceptibility of the chiral charge.

The interesting reason for writing the integrated form of the chiral anomaly equation in terms of the time derivative of the magnetic helicity is that it expresses a
conservation law. In the absence of processes that change the fermion chirality, i.e. when $\Gamma_f =0$, then Eq.~(\ref{anomal-full}) expresses that  the quantity ${\cal I} \equiv n_5 + \frac{C \alpha}{\pi} {\cal H}$ is
conserved. This means that an initial fermion chiral  imbalance can be converted into magnetic helicity, but also that a magnetic field with helicity might be
converted into a chiral fermion imbalance.  The last mechanism has been recognized in cosmological contexts~\cite{Giovannini:1997eg}, but we point out that it is a generic feature
of chiral systems that has been unnoticed in many studies of the CME, where a time-independent value of $\mu_5$ is generically considered.

Note that if $\Gamma_f =0$ the chiral anomaly equation allows one to make an estimate of the maximum magnetic helicity density that might
be created in the presence of an initial fermion chiral imbalance as 
\be
{\cal H}_{\rm max} \sim \frac{\pi}{C \alpha} n_5 (t=0) \ .
\ee
In practice, for the cases we will consider, this maximal value is
not always  achieved, as the magnetic field dynamics enters into a dissipative regime that does not allow for a sustained growth of the magnetic helicity.

In the remaining part of the manuscript we will study in more detail the anomalous Maxwell equations and consider how the helicity might be transferred from the fermions
to the gauge fields and vice versa at time scales $t \ll 1/\Gamma_f$.
A different solution to the chiral anomaly equation has been considered at large times,  
in the regime  $ t \gg 1/\Gamma_f$~\cite{Son:2012bg}, and in the presence of external parallel electric and magnetic fields. Such a solution
allows one to find a peculiar magnetic field dependence of the
electrical conductivity tensor that has been observed in Dirac semimetals~\cite{Li:2014bha}. In this work we will not
consider the presence of fixed background external fields, and rather concentrate
on the evolution of the chiral plasmas at shorter time scales.


\subsection{Solving  the AMEs in Fourier space}
\label{sec:FAME}

It is convenient to solve the AMEs.~({\ref{eq-Bfield}--\ref{anomal-full}) in Fourier space. After defining the orthonormal set of vector polarization vectors 
$({\bf e}_+,{\bf e}_-,{\bf \hat{k}})$ describing circular polarized waves~\cite{Jackson} we define the Fourier mode associated to the vector gauge field
potential
\be {\bf A}_{\bf k} = A_{\bf k}^+ {\bf e}_+ + A^-_{\bf k} {\bf e}_- + A_{\bf k}^k {\bf \hat{k}} \ . \ee
If the vector potential is real then
\be (A_{\bf k}^{\pm})^* = -A^\pm_{-{\bf k}} \ , \ee
and the magnetic field in Fourier space is then expressed as
\be {\bf B}_{\bf k}= - i {\bf k} \times {\bf A}_{\bf k} = B_{\bf k}^+ {\bf e}_+ + B_{\bf k}^- {\bf e}_- = - k (A_{\bf k}^+ {\bf e}_+ - A^-_{\bf k} {\bf e}_-) \ . \ee
It is possible to express the magnetic helicity density in terms of the circular polarized components of the magnetic field as
\be \label{eq:defhelicity}
{\cal H} = \frac 1V \int \frac{d \Omega_{\bf k}}{4 \pi} \int dk \frac{k}{2 \pi^2} \left( |B_{\bf k}^+ |^2 - |B_{\bf k}^- |^2  
\right) \ ,
\ee
where $d\Omega_{\bf k}$ is the solid angle element, not to be confused with the Berry curvature.

Written in this form, we observe that the magnetic helicity  gives account of an existing asymmetry in the magnetic field spectrum, measuring the difference
between its left- ($+$) and right-handed  ($-$) polarization components. Note also that when ${\cal H}$ is expressed in this form one clearly observes that the
 chiral anomaly equation (\ref{basicAnoeq}) relates the imbalance between left- and right-handed helicity components of both fermions and magnetic fields in a quite parallel way.
 
We note that the magnetic energy density can be written as
\be
\rho_B =  \frac 1V \int \frac{d \Omega_{\bf k}}{4 \pi} \int dk \frac{k^2}{(2 \pi)^2} \left( |B_{\bf k}^+ |^2  + |B_{\bf k}^- |^2
\right)  \ .
\ee
Using circular components, the coupled equations (\ref{eq-Bfield}) and (\ref{anomal-full}) reduce to
\ba
\label{AMW-1}
\frac{1}{\sigma}  \frac{\pa^2 B^\pm_{\bf k}}{\pa t^2} +  \frac{\pa B^\pm_{\bf k}}{\pa t} =- \left(\frac{1}{\sigma} k^2  \mp \frac{C \alpha  \mu_5 k}{\pi \sigma} \right)
 B^\pm_{\bf k} \ , \\
 \label{AMW-2}
 \frac{d n_5}{dt} =  - \frac{  C \alpha}{V}  \int \frac{d \Omega_{\bf k}}{4 \pi } \int dk \frac{k}{2 \pi^2} \frac{d}{dt}\left( |B_{\bf k}^+ |^2 - |B_{\bf k}^- |^2 \right) 
 - \Gamma_f n_5  \ .
 \ea

We see that the main effect of the existence of a fermion chiral imbalance is that the left and right circular polarized components of the magnetic field follow a different dynamical evolution.
It is easy to understand that in the presence of an axial chemical potential a nonvanishing magnetic helicity soon arises as a consequence of this wave asymmetry.

In the remaining part of this section we will work in the limit $\mu_5 \ll T$ and use Eq.~(\ref{eq-mu5}) with a $c_\Delta$ not depending on $\mu_5$. Hence, we can obtain simple solutions
and estimates for the dynamical evolution of the chiral imbalance and magnetic helicity.  However, in all our numerical examples we will consider the most general scenario, that may allow for larger values of the axial chemical potential.

\subsection{Solutions for small frequencies (or wave numbers)\label{sec:smallk}}

For small frequencies the coupled system  of equations described in Eqs.~(\ref{AMW-1}-\ref{AMW-2})
 greatly simplifies, and exact analytical solutions can be found. In this case, 
the equations have already been discussed in the high-$T$ limit (where $c_\Delta \propto \frac {1}{T^2}$ does not depend on $\mu_5$)
to study the magnetic field dynamics in cosmological contexts~\cite{Boyarsky:2011uy,Tashiro:2012mf} ---when one is interested in the behavior of the system at large distances.
These references focused on the appearance of a chiral magnetic instability, that might explain the existence of large cosmological magnetic fields.
Here, we review the form of the solutions, and explain why a chiral magnetic instability occurs.  In addition, we also present other solutions without instability
that show a growth of the magnetic helicity as a consequence of the chiral fermion imbalance. Further, we show that an initial large magnetic helicity might create a chiral fermion asymmetry.

Let us assume that the system is isotropic. If we neglect the second-order time derivative 
in Eq.~(\ref{AMW-1}) by assuming frequencies much smaller than the EM conductivity  $\omega \ll \sigma$, 
then the solution of this equation reads
\be
\label{SolAMW-1}
B^\pm_{\bf k} = B^\pm_{{\bf k},0}  \exp{\left[ -\frac{ (k^2 \mp K_p k)}{\sigma}t \right]} \ ,
\ee
where $ B^\pm_{{\bf k},0}$ is the initial condition  for the magnetic field, and we have defined
\be
K_p \equiv \frac{1}{t} \int^t_0  d\tau \frac{C \alpha \mu_5(\tau)}{\pi} \ .
\ee
We note that when $\mu_5$ is time independent then $K_p = \frac{C \alpha \mu_5}{\pi}$. 

The value of $\mu_5$ is then found by plugging the solution for the magnetic field above in Eq.~(\ref{AMW-2})
and solving the resulting integro-differential equation.  The solution can be expressed as
\be
\label{sol-mu5}
 \mu_5(t) = \mu_{5,0} \exp \left( - \Gamma_f t \right) - c_\Delta C \alpha [ {\cal H}(t)-{\cal H}(0) ] \ ,
 \ee 
and depends on the initial values of the axial chemical potential $\mu_{5,0} \equiv \mu_5(t=0)$ and magnetic helicity ${\cal H}(0)$, and on 
the chirality-flipping rate $\Gamma_f$.

We will now review, in a  general setting, how the chiral magnetic instability arises.
This was first discussed in Refs.~\cite{Boyarsky:2011uy,Tashiro:2012mf}, assuming an initial nonvanishing small
 value of the magnetic helicity, and in the approximation $\Gamma_f \approx 0$. 
Let us first assume that the magnetic field configurations are maximal helical 
i.e. $B^-_{{\bf k},0}=0$. Then, the system of equations greatly simplifies. 
Further, if one assumes a monochromatic magnetic helicity such that
\be \label{eq:monocrom}
 |B_{{\bf k},0}^+ |^2 = |B_0^+|^2  \delta(k -k_0) \ , 
 \ee
then the helicity at time $t$ can be expressed as 
\be \label{eq:maghel}
{\cal H} (t) = h_0 \exp \left[ \frac{2k_0}{\sigma} \left(  K_p - k_0 \right) t \right] \ ,  
\ee
where $h_0 = \frac{k_0}{2 \pi^2 V}  |B_0^+|^2$
and 
\be   \frac{d\mu_5}{dt} =
- \frac{2 c_\Delta C \alpha k_0}{\sigma } \left( \frac{C \alpha}{\pi} \mu_5 - k_0 \right) {\cal H}(t) \ . \ee

Then, one sees that for $k_0 > K_p$ the magnetic helicity decays exponentially, as the Ohmic dissipation dominates its large time behavior. However,
 if $k_0 <  K_p$ the magnetic helicity grows exponentially, signaling the presence of an instability.
The chiral imbalance then evolves until it reaches a tracking solution for $\mu_5$,
\be \label{eq:mutrack} \mu_{5,tr} = \frac{\pi k_0}{C \alpha} \ , \ee
at which the chiral imbalance remains constant in time. In this situation the growth of the magnetic helicity is exactly compensated by its decay due to the Ohmic dissipation. Therefore, as
${\cal H}$ remains constant, $\mu_5$ does not change in time.

The time scale where the instability starts to become effective can be estimated from the solution in Eq.~(\ref{eq:maghel}) as
\be
\label{time-hinstab}
t_{{\rm inst},h} = \frac{\sigma}{2k_0 (K_p-k_0)} \approx \frac{ \pi \sigma}{2k_0 C \alpha ( \mu_{5,0}-\mu_{5,tr})} \ ,
\ee
and it clearly depends on how far the initial value of the chemical potential is from the tracking solution.

 Looking at the equation for $\mu_5$, the characteristic time scale for the decrease of $\mu_5$ has a different
value
\be
t_{\mu_5} = \frac{\pi\sigma}{2 k_0 C^2  \alpha^2 c_\Delta \alpha {\cal H} }
 \approx \frac{\pi \sigma }{2 k_0 C^2  \alpha^2 c_\Delta  h_0} \ ,
\ee
which also depends on the initial value of the magnetic helicity density.

We show a particular example of the chiral magnetic instability in Fig.~\ref{fig:case1Inf}.
\begin{figure}[h]
\begin{center}
\includegraphics[scale=0.35]{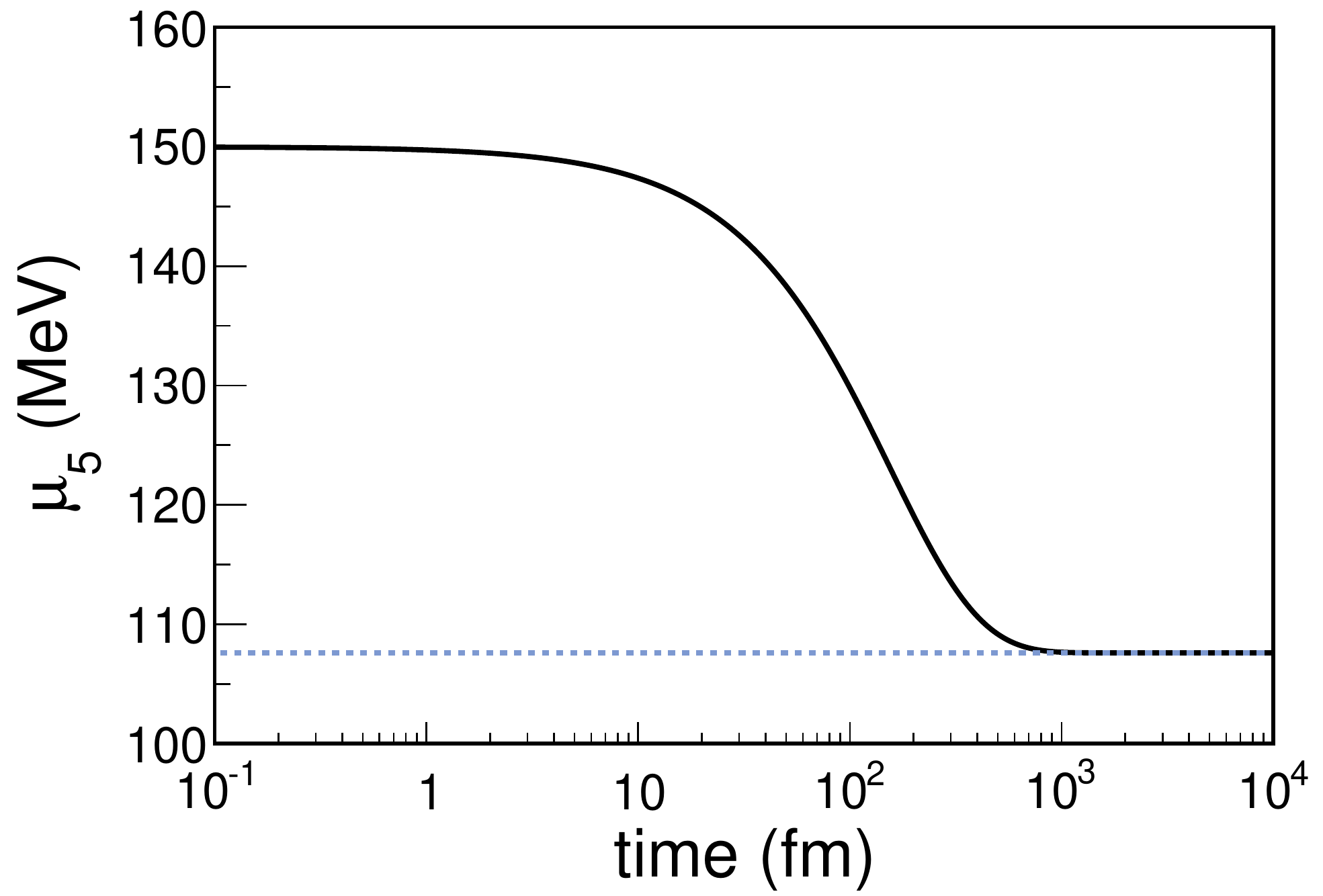}
\includegraphics[scale=0.35]{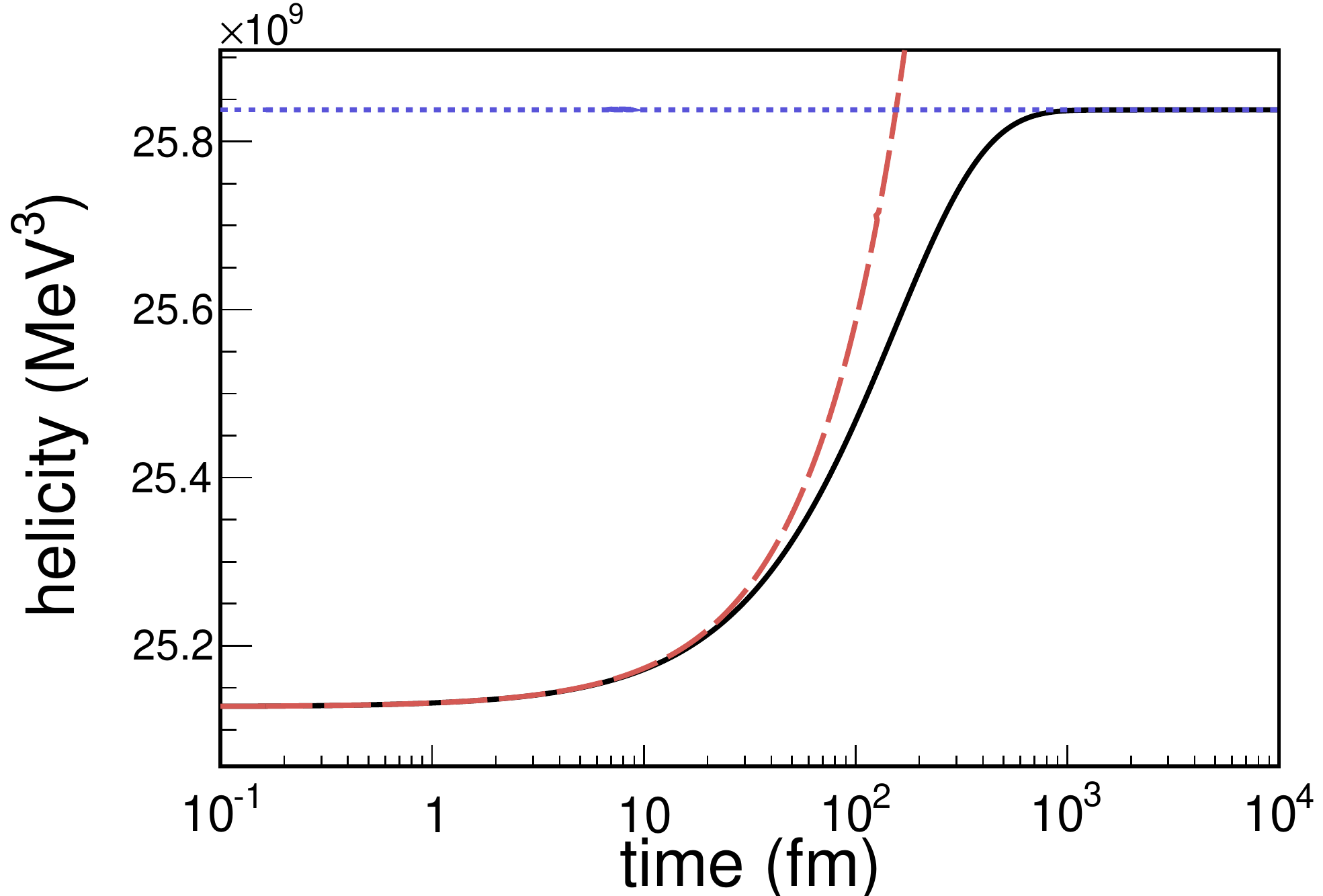}
\end{center} 
\caption{\label{fig:case1Inf} Chiral magnetic instability.
We use the parameters $T=225$ MeV, $\sigma=0.0244T$, and consider a sphere of radius $R=10^3$ fm, $k_0=0.5$ MeV, $\mu_{5,0}=150$ MeV, $\Gamma_f=0$, and
initial magnetic field as specified in the text. {\bf Left panel:} Chiral imbalance as a function of time (black solid line) and tracking solution 
given by Eq.~(\ref{eq:mutrack}) (blue dotted line). {\bf Right panel:} Magnetic helicity density as a function of time (black solid line), 
analytic solution at early times from Eq.~(\ref{eq:shorth}) (red dashed line), and maximum value for the magnetic
helicity density as given by Eq.~(\ref{eq:maxh}) (blue dotted line).}
\end{figure}
In this realization we set an initial $\mu_{5,0}=150$ MeV and use maximal helical magnetic fields (with $B^-_{{\bf k},0}=0$). 
We make the arbitrary choice $|B^+_{{\bf k},0}|^2 \sim k_0 \frac{|e {\bf B}_0 |^2 V^2}{ 4 \pi \alpha} $, where 
$|{\bf B}_{0}|$ is the magnitude of the initial constant magnetic field in configuration space, that we take as $|e{\bf B}_0|=m^2_\pi$, 
where $m_\pi = 135$ MeV is the pion mass. The other relevant parameters to determine the dynamical evolution 
are given in the caption of Fig.~\ref{fig:case1Inf}.

In the left panel of this figure, we observe the decrease of the initial chiral imbalance according to the AMEs with $k_0 < K_p$.
The dynamics is frozen when the chiral imbalance reaches the tracking solution in Eq.~(\ref{eq:mutrack}). In blue dotted line 
we show the value of this estimate.

In the right panel of Fig.~\ref{fig:case1Inf}, we present the magnetic helicity density (called simply ``helicity'' in
the figures) as a function of time. In black solid line we plot the numerical solution as obtained from our code.
At initial times ($t \ll \sigma/ 2 k_0^2$) an analytical estimate of the magnetic helicity can be obtained from Eq.~(\ref{eq:maghel})
\be \label{eq:shorth}
{\cal H}  = h_0 \left[ 1 + \frac{2k_0 C \alpha}{\sigma \pi} \left( \mu_{5,0}- \mu_{5,\textrm{tr}} \right) t
\right] \ .
\ee
This linear behavior is plotted using a red dashed line in Fig.~\ref{fig:case1Inf}. 
When the tracking solution for the chiral imbalance is reached, the magnetic helicity density remains constant. The maximum 
value of the magnetic helicity density can be determined from the anomaly equation~(\ref{anomal-full}) with $\Gamma_f=0$. It reads
\be \label{eq:maxh}
{\cal H}_{\textrm{max}} = h_0 + \frac{\pi}{C\alpha} \left( n_{5,0}- n_{5,\textrm{tr}} \right) \ ,
\ee
where $n_{5,\textrm{tr}}$ and $n_{5,0}$ are the tracking and initial value of the chiral number density, respectively.
The value for ${\cal H}_{\textrm{max}}$ is plotted in the dotted blue line in the right panel of Fig.~\ref{fig:case1Inf}, providing 
an additional check of our numerical code.

In the presence of several Fourier modes, the authors of Ref.~\cite{Boyarsky:2011uy} realized that there is an interesting inverse cascade phenomenon, 
where the magnetic helicity is transferred from the highest to the lowest modes, while the fermion chiral imbalance is washed out if one includes a small
and nonvanishing value of $\Gamma_f$. For a particular consideration of this case, we refer the reader to Ref.~\cite{Boyarsky:2011uy}.

We consider now a different situation, where the magnetic helicity might grow without the existence of an instability.
As before, let us first consider a monochromatic initial magnetic helicity in Fourier space, such as
\be
\label{monoB}
 |B_{{\bf k},0}^\pm |^2 = |B_0^\pm|^2  \delta(k -k_0) \ ,
 \ee
but with $|B_0^+|^2 =|B_0^-|^2 \equiv |{\tilde B}_0|^2$, so that one has an initial vanishing magnetic
helicity ${\cal H}(0)=0$.  
From the explicit solution of the dynamical evolution of the magnetic field, one can write the magnetic helicity density at time $t$ as
 \be
 \label{solH-zeroinitial}
{\cal H} (t) = 2 {\tilde h}_0 \exp{\left[ -\frac{2 k_0^2}{\sigma}t \right]} \sinh{\left(\frac{2 k_0 K_p}{\sigma}t \right)} \ ,
\ee
where ${\tilde h}_0 = \frac{|{\tilde B}_0|^2 k_0}{2 \pi^2 V}$. In the presence of a chiral imbalance, the magnetic
helicity grows at short times, although at large times it decays exponentially, due to
the Ohmic dissipation. More specifically, for $t \ll \frac{\sigma}{2 k^2_0}$ 
\be 
\label{shortt-h}
{\cal H}(t) \approx \frac{4 \tilde{h}_0 k_0 C \alpha \mu_{5,0} }{\pi \sigma} t \ .
\ee

This case clearly differs from the one discussed previously, as in the presence of the chiral instability the growth is exponential, and it only stops  when $\mu_5$ 
 reaches the tracking solution. In this case, the maximal  growth of ${\cal H}$ can be estimated  as follows: the Ohmic dissipation takes effect at times $t \sim \frac{\sigma}{2 k^2_0}$.
By this time, the magnetic helicity will have reached the approximated value 
\be
{\cal H} \left( t= \frac{\sigma}{2 k^2_0}\right) \simeq {\tilde h}_0 \sinh{\left(\frac{K_p}{k_0} \right)} \approx {\tilde h}_0 \sinh{\left(\frac{C\alpha \mu_{5,0}}{ \pi k_0} \right)} \ .
\ee
The growth of the magnetic helicity will depend on both the initial values of the magnetic field and chiral imbalance,
but also on the value of $k_0$. The growth can be very significant if $k_0 \ll \frac{C \alpha \mu_{5,0}}{\pi}$.

Quite similarly, one can obtain the time evolution of the magnetic energy density. One finds
\be
\rho_B(t) = \rho_B (0) \exp{\left[ -\frac{2 k_0^2}{\sigma}t \right]} \cosh{\left(\frac{2 k_0 K_p}{\sigma}t \right)} \ ,
\ee
and its maximum growth ---before the Ohmic dissipation takes effect--- can be estimated as
\be
 \rho_B (0)  \cosh{\left(\frac{K_p}{k_0}\right)} \approx   \rho_B (0)  \cosh{\left(\frac{C\alpha \mu_{5,0}}{ \pi k_0}\right)} \ .
\ee 

Now, it is easy to realize that the increase of the magnetic helicity comes together with a lowering of the axial chemical potential, as dictated by Eq.~(\ref{sol-mu5}).
In the presence of a small nonvanishing helicity-flipping rate $\Gamma_f$, $\mu_5$ would also tend to vanish at later times. However if
 $\Gamma_f \approx 0$, the chiral anomaly (which expresses the conservation of the helicity of fermions and gauge fields) dictates that the fermion chiral imbalance should grow when
the magnetic helicity enters the regime where the Ohmic dissipation is effective.
This is illustrated in Fig.~\ref{fig:case2Inf} for a plasma of fixed spherical volume, where we study both the fermion chiral imbalance and magnetic helicity density for a monochromatic field, in the regime
$k \ll \sigma$. In all the examples in this section we assume  $n_5 \approx \frac 32 \mu_5 T^2$, and make the arbitrary choice  $|\tilde{B}_{0}|^2 \sim k_0 \frac{|e {\bf B}_0 |^2 V^2}{ 4 \pi \alpha} $, where 
$|{\bf B}_{0}|$ is the magnitude of the initial constant magnetic field in configuration space,  and we take  $|e{\bf B}_0|=m^2_\pi$, 
where $m_\pi = 135$ MeV is the pion mass.
The explicit values of the other relevant parameters are detailed in the figure's caption and are taken
from some of the conditions that are found in a QGP; see Sec.~\ref{Sec-QGP}.

In Fig.~\ref{fig:case2Inf} one observes a temporary creation of magnetic helicity due to the decrease of the initial chiral imbalance $\mu_{5,0}=10$ MeV.
We also plot in the red dashed line the estimate Eq.~(\ref{shortt-h}) that approximates the increase of the magnetic helicity 
at initial times.
 
\begin{figure}[h]
\begin{center}
\includegraphics[scale=0.35]{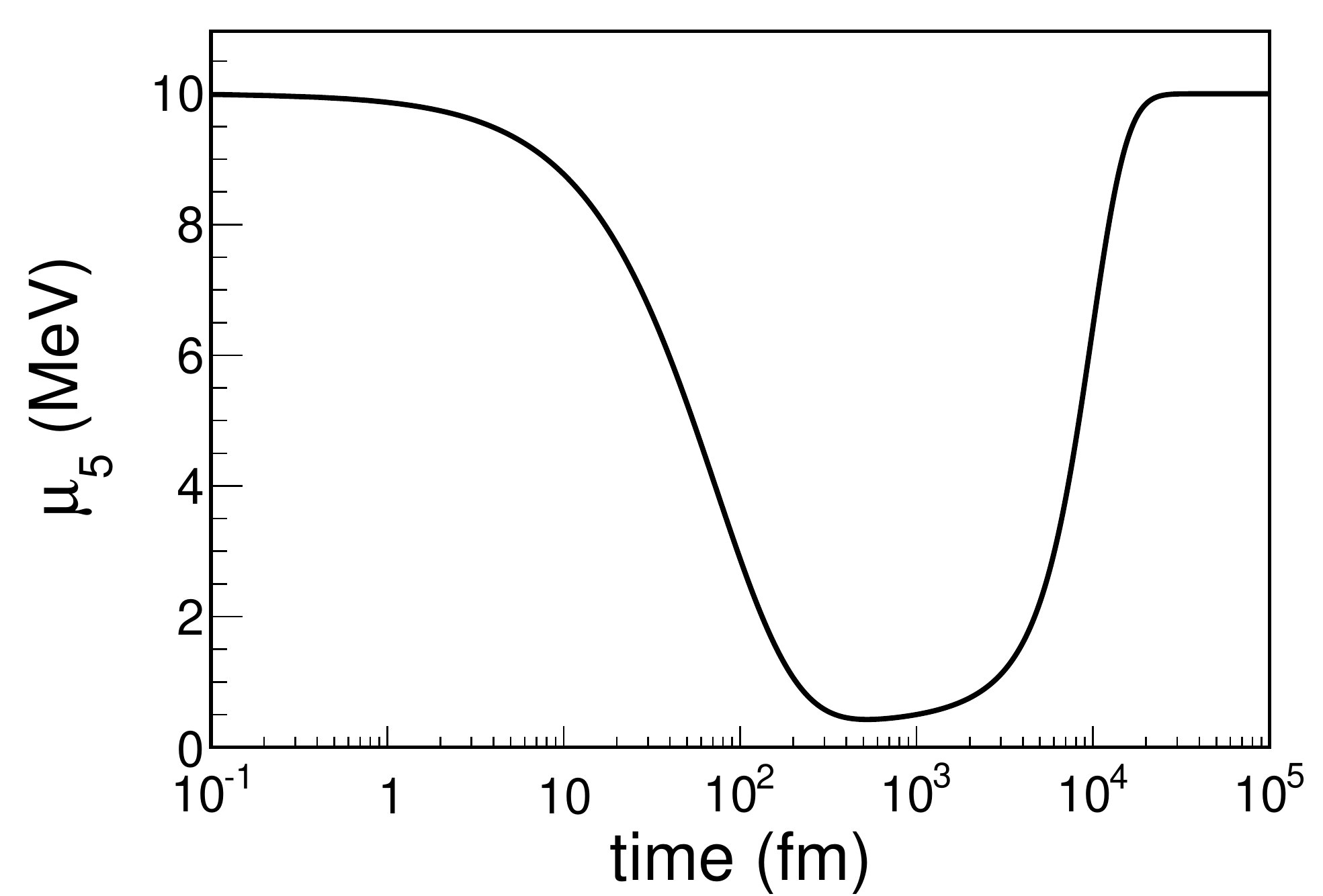}
\includegraphics[scale=0.35]{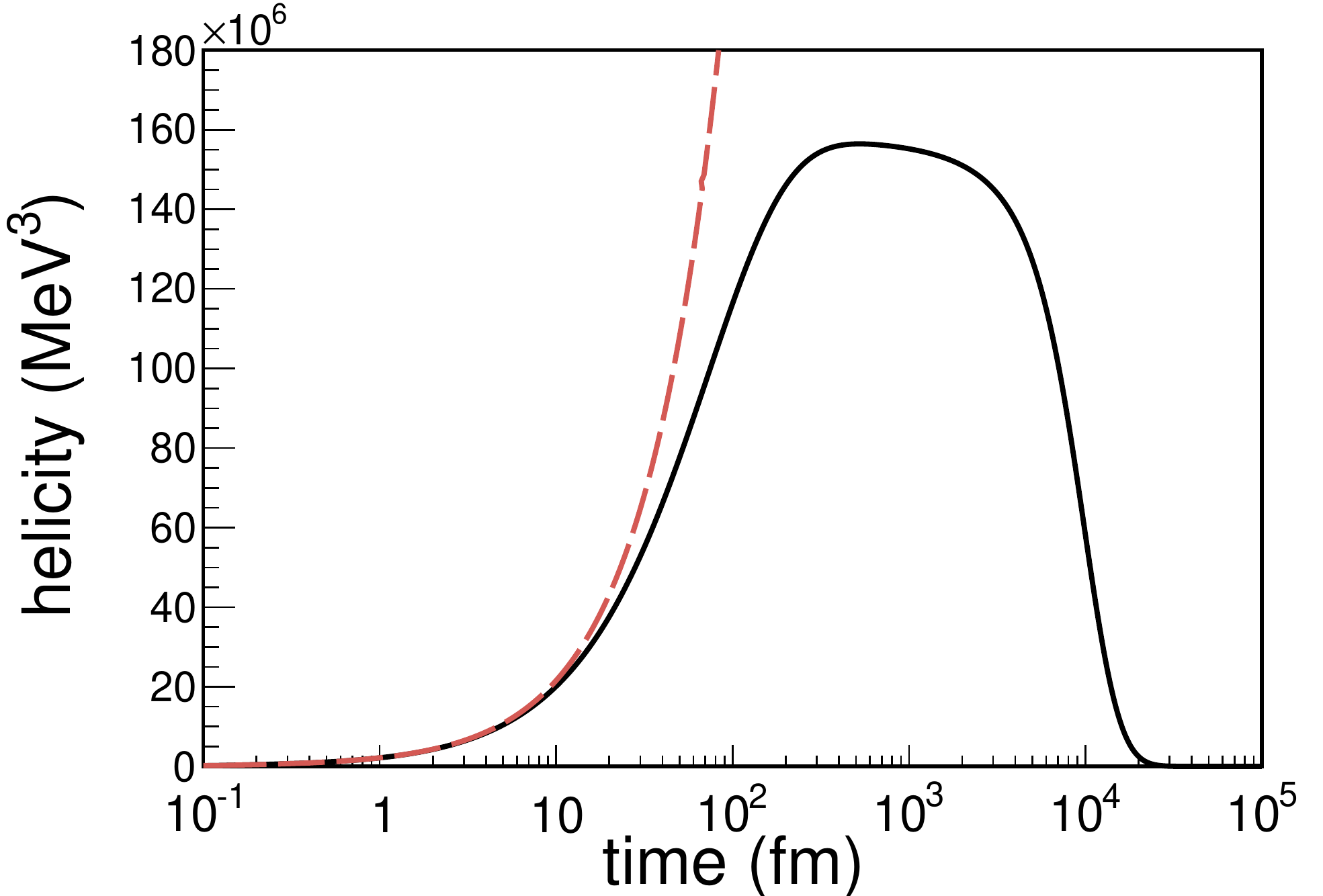}
\end{center} 
\caption{\label{fig:case2Inf} Temporary generation of magnetic helicity at the expenses of an initial fermion chiral imbalance. 
We use the parameters $T=225$ MeV, $\sigma=0.0244T$, and consider a sphere of radius $R=10^3$ fm, $k_0=0.5$ MeV, $\mu_{5,0}=10$ MeV, $\Gamma_f=0$, and
initial magnetic field as specified in the text. {\bf Left panel:} Chiral imbalance as a function of time. {\bf Right panel:}
Magnetic helicity density as a function of time (black solid line) and analytic solution at early times given by Eq.~(\ref{shortt-h})
(red dashed line).}
\end{figure}

Finally, let us describe the behavior of the system when several Fourier modes of the {\bf B} field are taken into account. If we now assume
\be
 |B_{{\bf k},0}^\pm |^2 =\sum_{i=1}^N |\tilde{B}_{0,i}|^2  \delta(k -k_i) \ ,
 \ee
where to ensure a vanishing initial magnetic helicity, we assume that $ |\tilde{B}_{0,i}|^2 = |B_{0,i}^+|^2=|B_{0,i}^-|^2 $. $N$ is the number of Fourier modes.
From the AME, all modes evolve independently and the total magnetic helicity density is the linear sum of partial helicities
\be 
{\cal H} (t) = \sum_{i=1}^N {\cal H}_i (t) = \sum_{i=1}^N  \ 2 \ \tilde{h}_{0,i} \exp \left[ -2 \frac{k_i^2}{\sigma} t \right] \ \sinh \left[ 2 \frac{K_p k_i}{\sigma} t \right] \ ,  
\ee
where $\tilde{h}_{0,i} = \frac{k_i}{2 \pi^2 V}  |\tilde{B}_{0,i}|^2$, where we further assume $|\tilde{B}_{0,i}|^2 \sim k_i \frac{|e {\bf B}_0 |^2 V^2}{ 4 \pi \alpha} $. 

Our analysis shows that the dynamical evolution of the modes with higher $k_i$ would be much faster and would grow more than those of lower values of $k_i$. At very short time scales, Eq.~(\ref{shortt-h}) tells us that
the growth of ${\cal H}_i$ depends on  the product $k_i {\tilde h}_{0,i}$, and ${\tilde h}_{0,i}$ which  depends on $k_i$. On the other hand, the helicity of the mode with higher $k_i$ decays faster, as this happens at $t \sim \frac{\sigma}{2 k_i^2}$.
All the modes are coupled by the equation of the anomaly
\be \frac{d\mu_5}{dt} = - \frac{2c_\Delta C \alpha}{\sigma} \sum_{i=1}^N \ k_i \left( \frac{C\alpha \mu_5}{\pi} - k_i \right) {\cal H}_i (t) \ . \ee

Notice that when the first mode dies out due to Ohmic dissipation, the modes with a smaller wave number still contribute
to the evolution of the chiral imbalance. Then, $\mu_5$ keeps evolving until the mode with smallest $k_i$ decays.
This is illustrated in Fig.~\ref{fig:case2polyInf}, where we have used three Dirac modes $k_i=\{0.2, 0.4, 0.6\}$ MeV plotted in 
red, blue and green lines, respectively, while the total magnetic helicity is plotted in black. Notice that according to our model the amplitude of each mode 
$\tilde{h}_{0,i}$ depends on the wave number. The remaining parameters are chosen as in Fig.~\ref{fig:case2Inf}.  We show the estimate of Eq.~(\ref{shortt-h}) for each individual mode in dashed lines.

\begin{figure}[h]
\begin{center}
\includegraphics[scale=0.35]{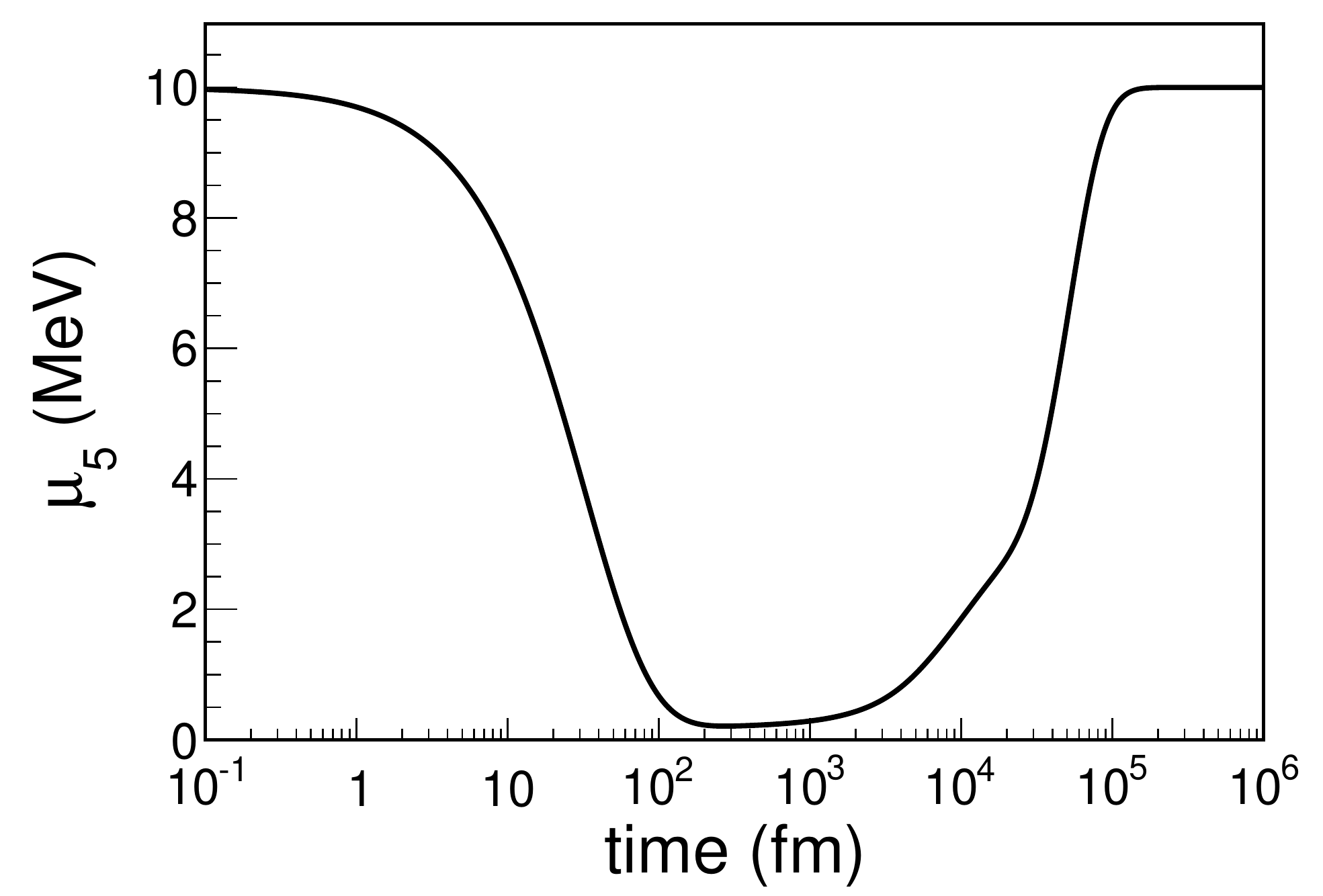}
\includegraphics[scale=0.35]{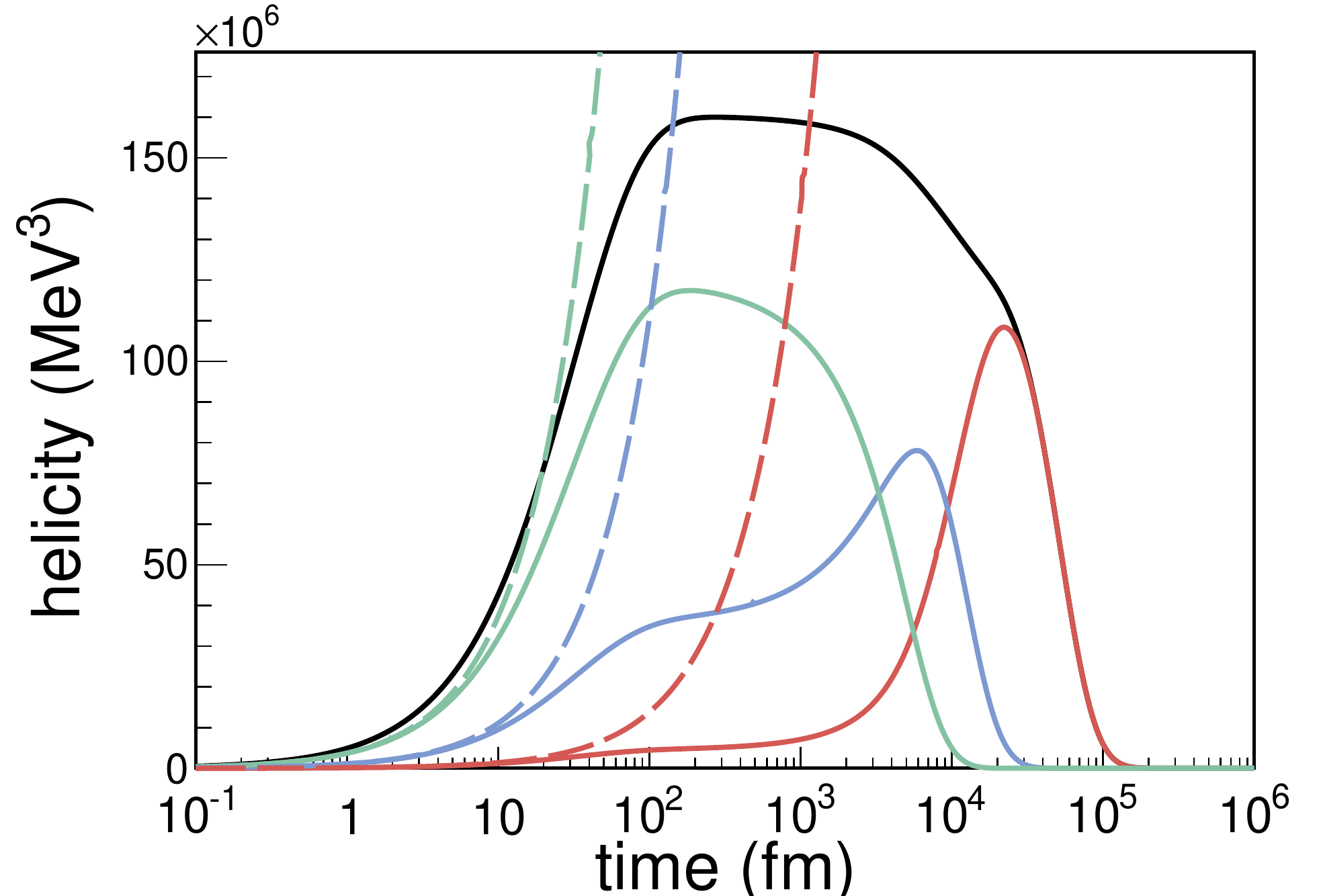}
\end{center} 
\caption{\label{fig:case2polyInf} Temporary generation of magnetic helicity at the expenses of an initial chiral imbalance for a polychromatic case. 
We use the parameters $T=225$ MeV, $\sigma=0.0244T$, and a sphere of radius $R=10^3$ fm, $k_i=${0.2 (red), 0.4 (blue), 0.6 (green)} MeV, $\mu_{5,0}=10$ MeV, $\Gamma_f=0$.
{\bf Left panel:} Chiral imbalance as a function of time. {\bf Right panel:} In solid lines we plot the total magnetic helicity density (black curve) and individual 
contributions to it for each of the three modes (red, blue and green lines) as a function of time. In dashed lines we include the 
estimates of Eq.~(\ref{shortt-h}) for early times.}
\end{figure}

Finally, let us mention that Eqs.~(\ref{AMW-1}-\ref{AMW-2}) also allow for a transfer of magnetic helicity density
to the fermions. 
Interestingly, the chiral anomaly equation suggests that whatever mechanism that may damp the magnetic helicity might be
a source of chiral fermion imbalance. For an electromagnetic conductor, this mechanism is the Ohmic dissipation.

A configuration with an initial large
magnetic helicity reduces its value, then creating chiral fermion  imbalance.
As an example, we present the solution for vanishing initial chiral imbalance but large magnetic helicity 
in Fig.~\ref{fig:case3Inf}. For simplicity we use a single Fourier mode at $k_0=0.5$ MeV. The reduction of the initial magnetic helicity 
and the generation of chiral imbalance is maintained until the tracking value $\mu_{5,\textrm{tr}}=\pi k_0/C\alpha \simeq 107$ MeV is reached.
At this point both variables are constant in time, unless there is an independent process that reduces the chiral fermion imbalance,
represented by a nonvanishing value of $\Gamma_f$. This last case is represented in dashed lines, where we solve the same equation with a $\Gamma_f=1/10^3$ fm$^{-1}$.
\begin{figure}[h]
\begin{center}
\includegraphics[scale=0.35]{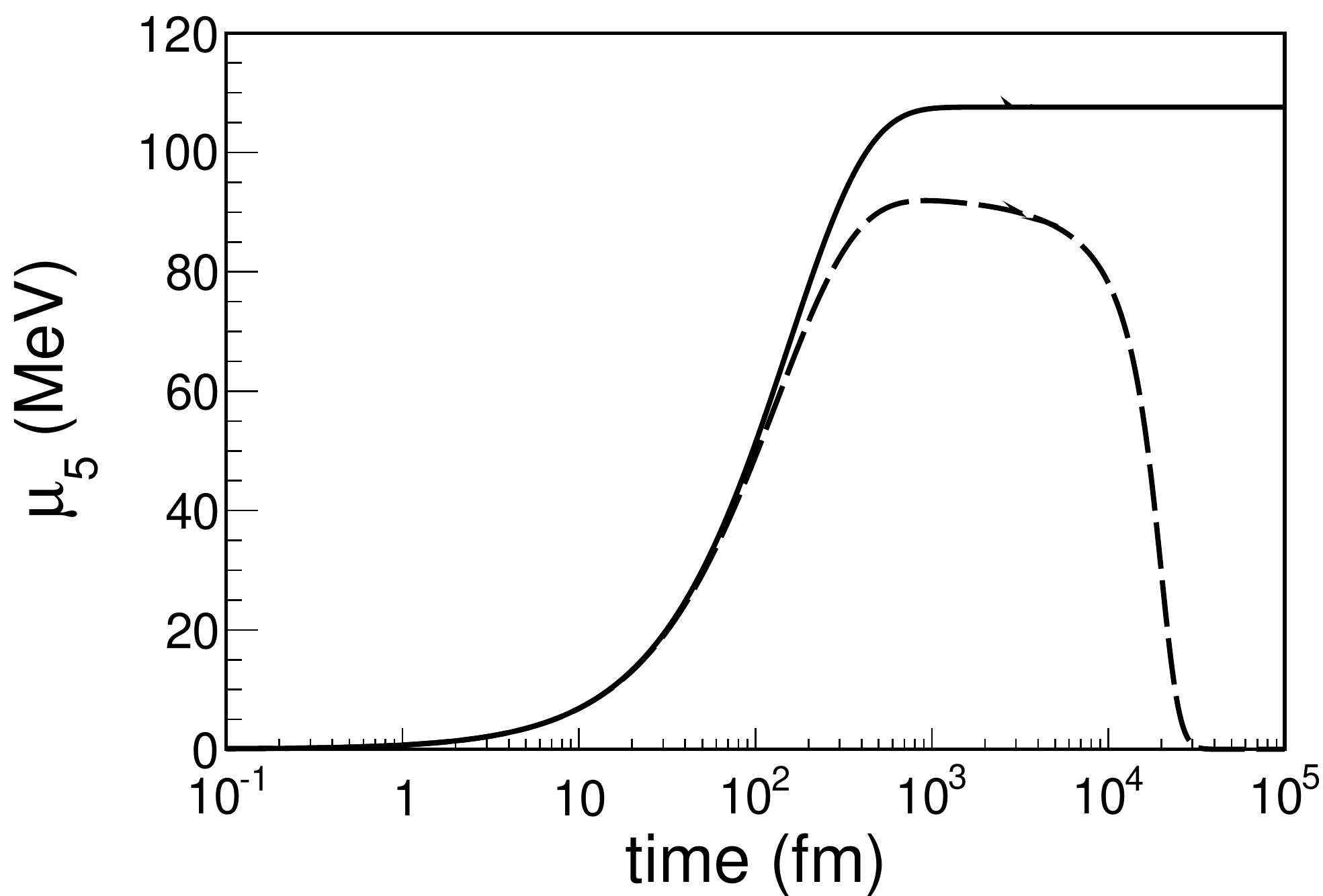}
\includegraphics[scale=0.35]{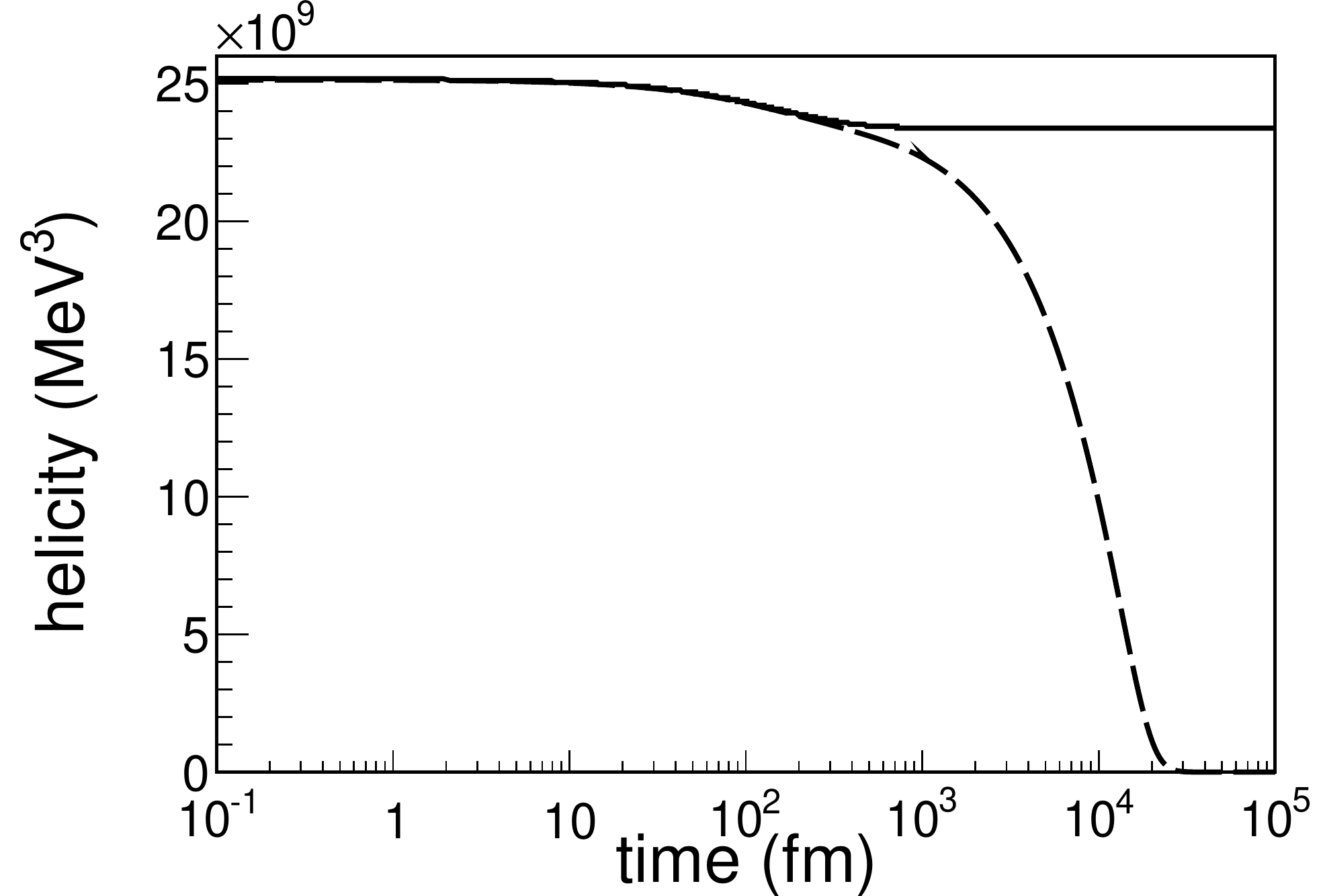}
\end{center} 
\caption{\label{fig:case3Inf} Generation of chiral imbalance due to the presence of an initial magnetic helicity.
We use the parameters $T=225$ MeV, $\sigma=0.0244T$, $R=10^3$ fm, $k_0=0.5$ MeV, $\mu_{5,0}=0$ MeV, $\Gamma_f=0$.
In dashed lines we plot the solutions to the AMEs when a $\Gamma_f=1/10^3$ fm$^{-1}$ is included.
{\bf Left panel:} Chiral imbalance as a function of time. {\bf Right panel:} Magnetic helicity density as a
function of time.}
\end{figure}

\subsection{Solutions for large frequencies (or wave numbers)}\label{sec:largek}

While we have seen that analytical solutions can be found for the coupled system of Eqs.~(\ref{AMW-1}-\ref{AMW-2}) when second-order time derivatives are
neglected, in the most general case this is not possible, and a numerical study is required. An analytical solution is only possible if one further assumes that  $\mu_5$ is constant. In such a case,
assuming  that the magnetic fields have a time dependence of the form $\sim e^{i \omega_{\pm} t}$, then the frequency is the solution of
\be
\omega_{\pm}^2 - i \sigma \omega_{\pm} - k^2 \pm \frac{C \alpha \mu_5}{\pi} = 0 \ .
\ee
For both left- and right-handed polarized components there are two solutions: 
\be
\omega_{\pm} = \frac{i \sigma}{2} \left( 1 \pm \sqrt{\phi_\pm} \right) \ , \qquad  \phi_{\pm} \equiv 1 - \frac{4 k}{\sigma^2} \left( k \mp \frac{C \alpha \mu_5}{\pi}  \right) \ .
\ee
If we impose as boundary conditions $B^{\pm}_{\bf k} (t=0) = B^{\pm}_{\bf k,0} $ and $\frac{\pa B^{\pm}_{\bf k}}{\pa t} \big|_{t=0}= 0$, then the solutions for both polarizations read
\ba
\label{generalSOL}
B^\pm_{\bf k} = \frac 12 B^\pm_{{\bf k},0} \left[ e^{- \frac{ \sigma t }{2} \left( 1 + \sqrt{\phi_\pm} \right) } \left( 1 - \frac{1}{\sqrt{\phi_\pm}}\right) +   e^{ - \frac{ \sigma t }{2} \left( 1 - \sqrt{\phi_\pm} \right) } \left( 1 + \frac{1}{\sqrt{\phi_\pm}}\right)  \right] \ .
\ea

It is possible to check that when $k \ll \sigma$, Eq.~(\ref{generalSOL}) agrees with Eq.~(\ref{SolAMW-1}), for a constant $\mu_5$. This condition guarantees that $\omega \ll \sigma$, which explains
why the higher time derivatives can be neglected in the study of the system.

First, we observe that for large wave numbers a  chiral magnetic instability can also occur. Let us consider that we deal with maximal helical fields, $B^-_{\bf k} =0$. Equation (\ref{generalSOL}) for $B^+_{\bf k}$ can describe an
exponentially growing solution whenever $\sqrt{\phi_+} > 1$.  This occurs for values of the wave number satisfying
$k < \frac{C \alpha \mu_5}{\pi}$ , and assuming $ k > \sigma$. We also find that at the value of the tracking solution given by Eq.~(\ref{eq:mutrack})
the magnetic field remains constant. The time scale when the
growth is effective is
\be
t_{\rm inst,h} \sim \frac{1}{\sigma (\sqrt{\phi_+}-1)} \ .
\ee
Note that for small wave numbers this occurs at a time scale given by Eq.~(\ref{time-hinstab}).

However, if we assume $k \gg \sigma$ and $k > \frac{C \alpha \mu_5}{\pi}$ then both $\phi_{\pm} < 0$, and the solutions for both left and right  circular waves become oscillatory,

\be
\label{oscillB}
B^\pm_{\bf k} = B^\pm_{{\bf k},0} e^{- \frac{ \sigma t }{2} } \left[  \cos{ \frac{\sigma \Delta_\pm t}{2}}  + \frac{1}{\Delta_\pm} \sin{ \frac{\sigma \Delta_\pm t}{2}}  \right] \ ,
\ee
where we have defined $\Delta_\pm \equiv  \sqrt{|\phi_\pm|}$. Then, we see that all the modes decay at time scales of order $t \sim 2/\sigma$, independent on the value of the wave number, although the oscillations at shorter time scales
depend on its value. With this form of the solution, the magnetic helicity should decay at time scales of the order $t \sim 1/\sigma$.

While we cannot present analytical forms of the solution if we assume a nonconstant $\mu_5$, the above solutions give us a hint of how the system behaves when we consider full time dependence of
$\mu_5$. In this last case, the coupled equations have to be studied numerically. Details on  how we have carried out such an analysis are given in Appendix~\ref{app:numerics}.

We consider different situations with a fermion chiral  imbalance for both monochromatic and polychromatic cases. In our numerical analysis we emulate the
Dirac delta functions of Eq.~(\ref{monoB}) with a very narrow Gaussian, the amplitude of which is chosen to provide a correct limit to the Dirac delta function, see Eq.~(\ref{eq:BGaussian}).

We assume a Gaussian profile for the magnetic field peaked at $k_0=100$ MeV and width $\kappa=3$ MeV, the amplitude is fixed as in the previous subsection.
The magnetic helicity is chosen to
be zero  by taking the same value for the left- and right-handed polarized components of the magnetic field. The initial chiral imbalance is taken
$\mu_{5,0}=100$ MeV. The results are shown in Fig.~\ref{fig:N1_k100_s3_m100_B1_h0}.
  \begin{figure}[ht]
 \begin{center}
 \includegraphics[scale=0.35]{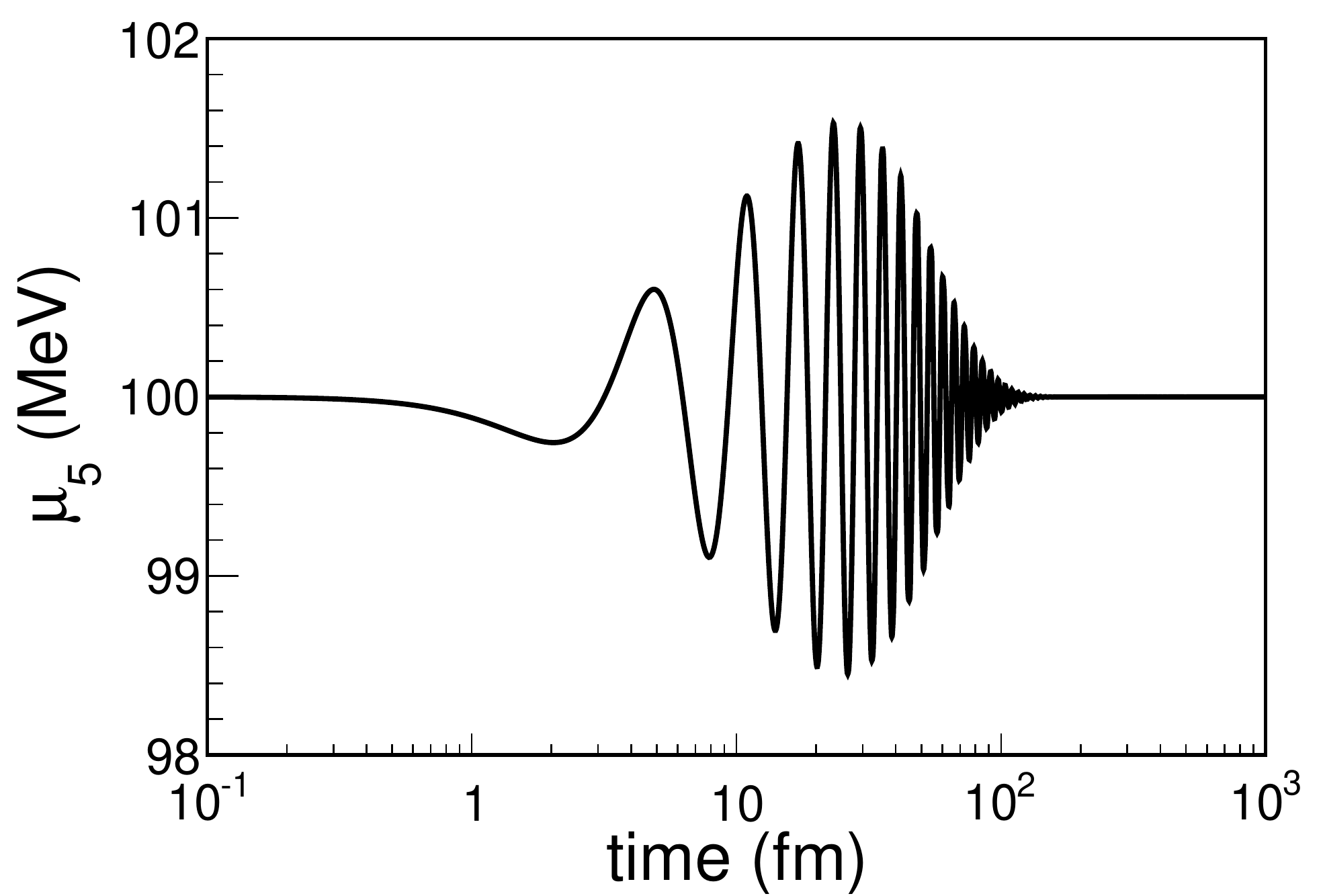}
 \includegraphics[scale=0.35]{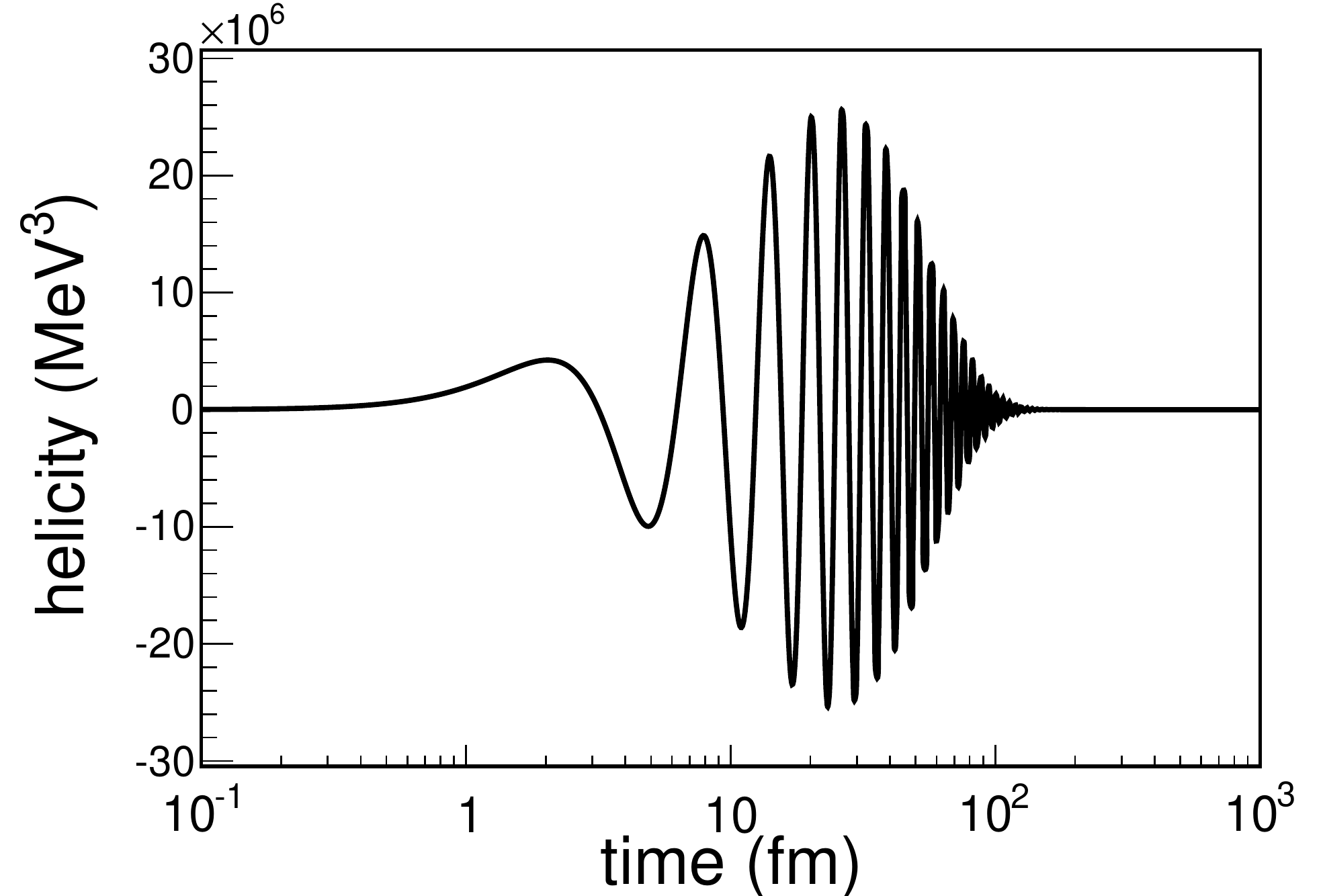}
 \end{center} 
 \caption{\label{fig:N1_k100_s3_m100_B1_h0} Generation of magnetic helicity due to an initial chiral imbalance for a monochromatic field with $k_0= 100$ MeV. As initial conditions, we set a chiral imbalance of $\mu_{5,0}=100$ MeV and zero magnetic helicity density. 
 We use the same values of $\sigma$ and $T$ that were chosen in previous figures,
but reduce the volume of the sphere and take $R = 10$ fm.}
 \end{figure}
Being close to a Dirac delta with $k_0 \gg \sigma$  and $k > C\alpha \mu_5 / \pi$, oscillatory solutions are found for $\mu_{5}$ and ${\cal H}$. These oscillations only disappear
at $t \gg 1/\sigma \simeq 40$ fm. On average, the trend coincides with the findings corresponding to the analytical solutions for constant $\mu_5$ displayed in Eq.~(\ref{oscillB}): the solutions are oscillatory and the initial chiral imbalance induces a temporary nonzero magnetic helicity 
until the latter decays at $t \simeq 1/\sigma$.

In Fig.~\ref{fig:N3_k100_200_300_s3_m100_B1_h0}  we consider the same situation for a polychromatic magnetic field, with
$N=3$ and wave numbers $k_i=100, 200, 300$ MeV, and with the same specific choice of
dependence on $k_i$ of the initial spectrum that was used to analyze the evolution for small wave numbers. There are interferences of the oscillations associated to every Fourier mode.
Such interferences depend, however, on the initial wave number spectrum of the magnetic fields.
  \begin{figure}[ht]
 \begin{center}
 \includegraphics[scale=0.35]{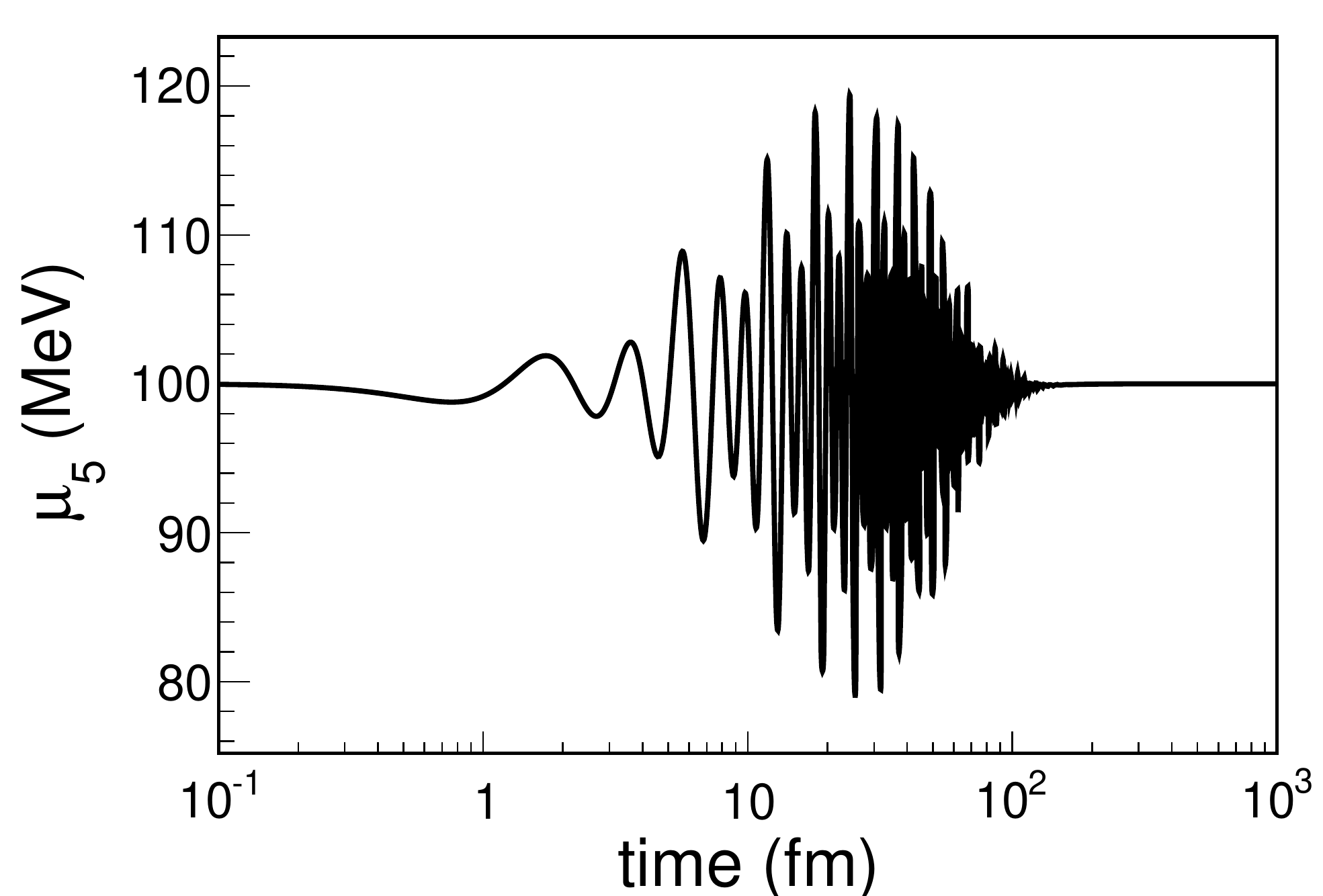}
 \includegraphics[scale=0.35]{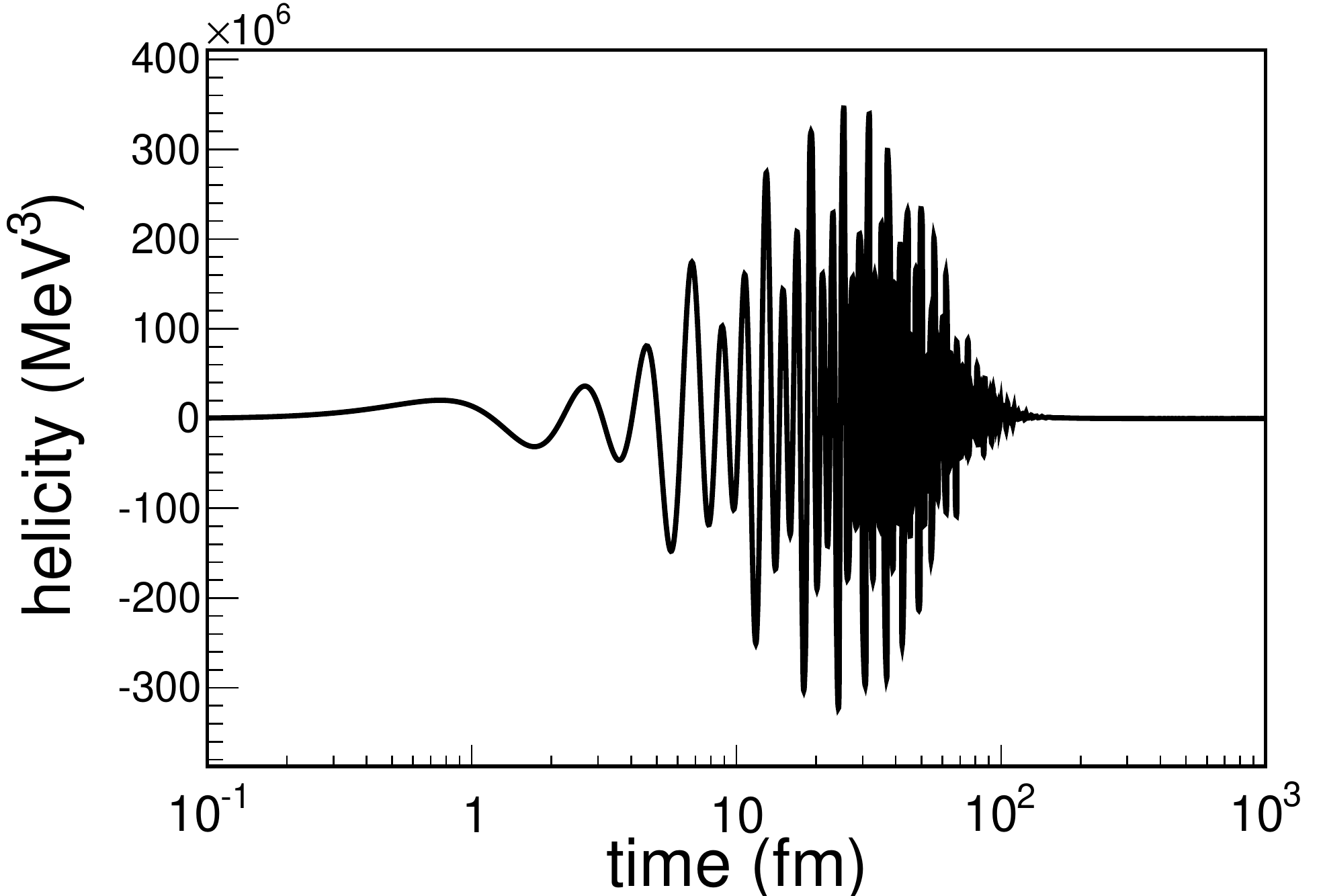}
 \end{center} 
 \caption{\label{fig:N3_k100_200_300_s3_m100_B1_h0}  Same plots as in Fig.~\ref{fig:N1_k100_s3_m100_B1_h0} but using a polychromatic field with 
wave numbers: $k_i=100, 200, 300$ MeV. Notice the interferences among the three modes.}
 \end{figure}

 Finally, we consider the case in which a chiral imbalance is generated due to the presence of an initial magnetic helicity.
We use the single Gaussian spectrum centered at $k_0=100$ MeV and $\kappa=3$ MeV (monochromatic case). The initial amplitude of the magnetic field is taken
as in the previous examples, but with a vanishing right-handed component. In Fig.~\ref{fig:N1_k100_s3_m0_B1_h1} we see how the initial magnetic helicity density 
generates a chiral imbalance which oscillates in time. Again, the helicity decays in times of the order of $t\sim 1/\sigma=40$ fm, creating a
stable chiral imbalance that might eventually vanish by the effect of some $\Gamma_f$.
  \begin{figure}[ht]
 \begin{center}
 \includegraphics[scale=0.35]{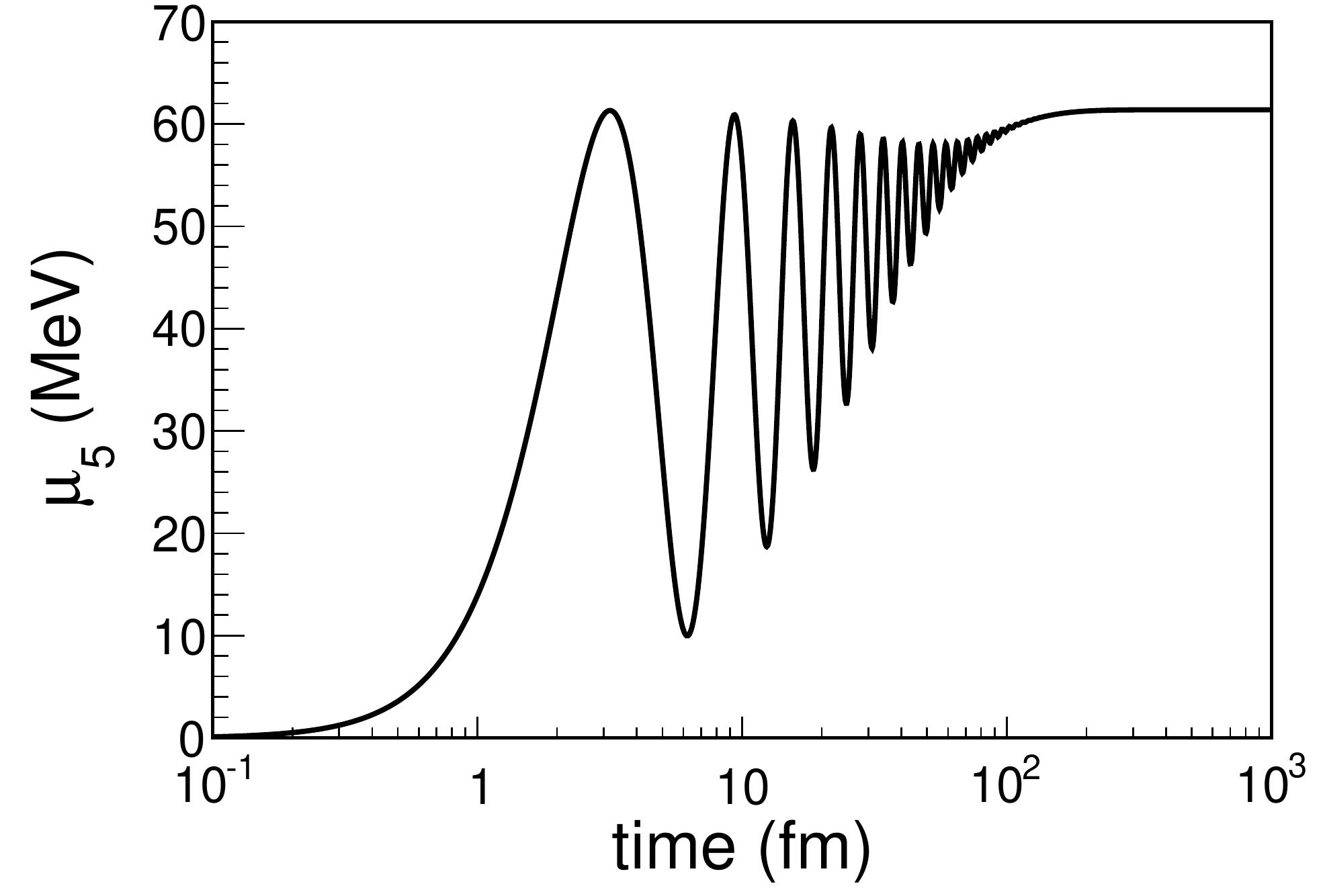}
 \includegraphics[scale=0.35]{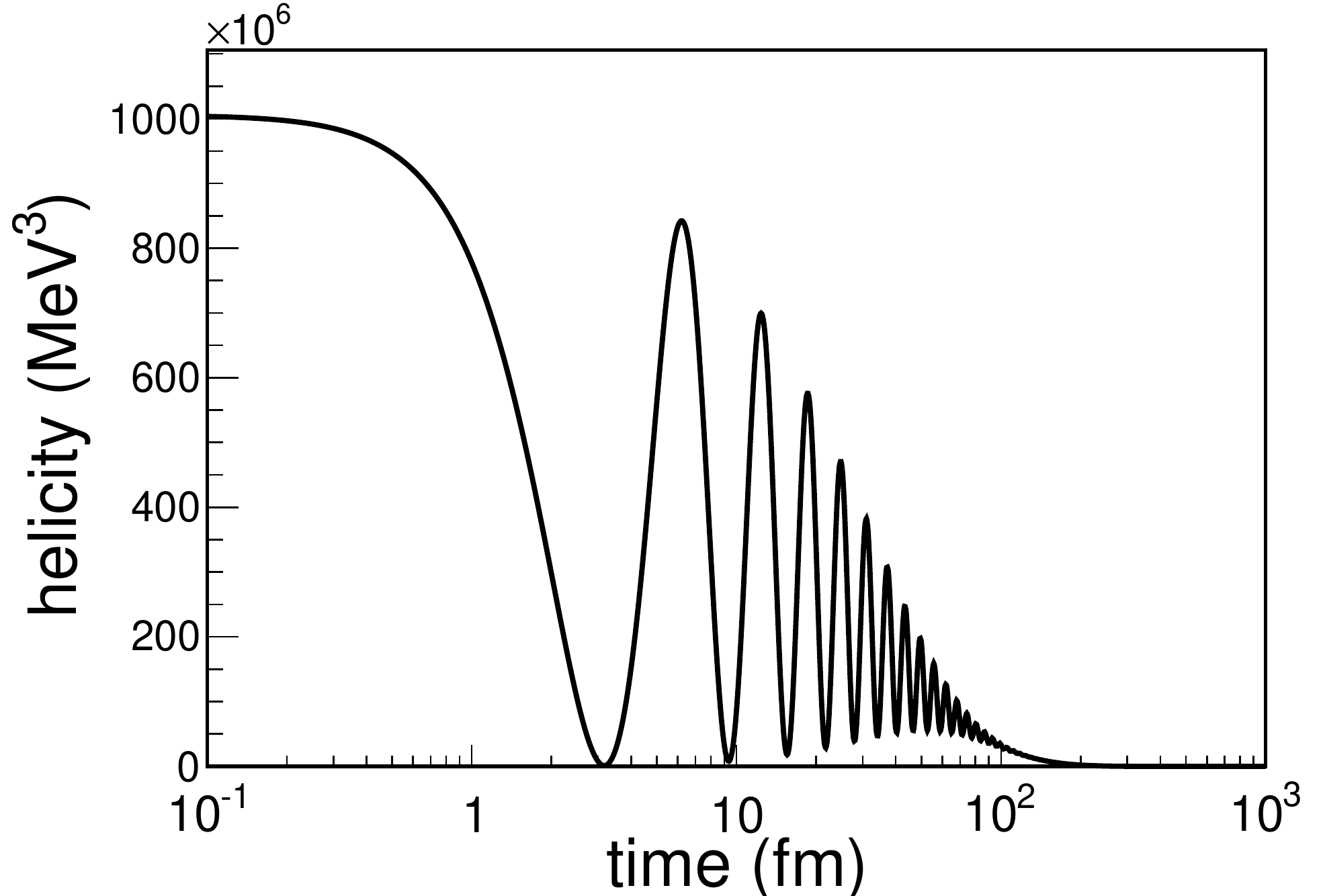}
 \end{center} 
 \caption{\label{fig:N1_k100_s3_m0_B1_h1} Generation of chiral imbalance due to an initial magnetic helicity in the system.
We set an initial vanishing chiral imbalance and a nonzero initial magnetic helicity density by choosing a left-handed-polarized magnetic field
and a vanishing right-handed component.
The profile of the initial magnetic field in Fourier space is a Gaussian centered at $k_0=100$ and width $\kappa=3$ MeV.}
 \end{figure}

Summarizing, we have  analyzed the time evolution for monochromatic and polychromatic (and unrealistic) magnetic fields, and observe that the transfer
of helicity from fermions to magnetic fields behaves very differently for large or small wave numbers. A much more realistic case should consider
the full Fourier mode spectrum, and the final result would depend on the initial spectrum of magnetic fields. In the following section, we will show
the behavior of the helicity transfer when the initial spectrum is a wide Gaussian for the QGP conditions.

\section{Application to the QGP}
\label{Sec-QGP}

It has been claimed that the CME might have several phenomenological consequences in relativistic heavy-ion collisions~\cite{Kharzeev:2007jp,Fukushima:2008xe,Kharzeev:2010gr}.
In these experiments a chiral  imbalance between left- and right-handed quarks might arise as a result of an event-by-event fluctuation of the non-Abelian topological charge.
While averaged over all events, the topological charge should be zero, a gluon field configuration with nonvanishing winding number should be converted, thanks to the non-Abelian chiral anomaly, in a difference
in the populations of left- and right-handed quarks. Further, in  peripheral heavy-ion collisions, large values of the magnetic fields are expected to occur,  of the order of
$e |{\bf B}_0| \sim (1-10) m^2_\pi$~\cite{Skokov:2009qp}. It has been claimed that the CME should have an imprint on the charge asymmetry fluctuations detected in HICs~\cite{Kharzeev:2010gr}. Unfortunately, it is
not  easy to identify the occurrence of this quantum effect in these experiments, and different experimental signatures of the CME in HICs have to be studied.

Studying the dynamical evolution of the chiral magnetic effect in the QGP is much more involved than in the systems discussed in the previous part of this manuscript.
The main reason is due to the fact the chiral anomaly is also affected by the non-Abelian gauge fields. Thus, rather than Eq.~(\ref{chiral-anomaly})
the total axial current obeys~\cite{Peskin:1995ev}
\be
\partial_\mu j^\mu_A = \frac{ 2 C \alpha}{\pi}  {\bf E} \cdot {\bf B} + \frac{\alpha_s N_f}{\pi }  {\bf E}_a \cdot {\bf B}_a \ ,
\ee
where ${\bf E}_a$ and ${\bf B}_a$ are the chromoelectric and chromomagnetic fields, respectively, and $\alpha_s$ is the QCD
fine structure constant.  We have neglected above (and will do it in the following) corrections to the chiral anomaly equation due to nonvanishing values
of the quark masses $m_q$, which might be entirely justified for high values of the temperature, if we are in the regime where $m_q \ll T$.

It is then clear that in the QGP the evolution of the quark chiral  imbalance  has to be studied together with both the Maxwell and Yang--Mills equations associated to the
system. We should note that it has been claimed that in the presence of a quark chiral imbalance, there should be also chiral chromomagnetic instabilities, very similar
to what happens in the Abelian sector~\cite{Akamatsu:2013pjd,Akamatsu:2014yza}. However, our claim is that this can only be so in the limit of very weak couplings,
$\alpha_s \ll 1$, and it is reasonable to expect that at large and also at moderate small values of $\alpha_s$, as those that are expected to occur in HICs, the nonlinearities
of the gluon dynamics would saturate the growth of the chromomagnetic fields. This is at least what happens in the case of Weibel instabilities in a non-Abelian plasma~\cite{Arnold:2005vb}.
This issue requires a further detailed study, that we will not address here.

 At the initial stages of the HIC, a nonvanishing value of the topological charge associated to the gluon fields might create
the quark chiral imbalance. At the very early stages of the collision, the chromoelectromagnetic fields are so strong that is reasonable  to ignore the effect
of the electromagnetic fields in the evolution of the quark chiral imbalance. In this situation the evolution of quark chirality might be affected by  the contribution of sphaleronlike transitions. QCD sphalerons allow to change the Chern--Simons number associated to the gluon fields, and
this change,  due to the chiral anomaly associated to the non-Abelian gauge fields,  leads to a variation of the quark chiral fermion number \cite{McLerran:1990de}.
In particular, in Ref.~\cite{Moore:2010jd}, and making use of a fluctuation-dissipation result, the time evolution of the chiral quark number due only to QCD sphaleron transitions is 
expressed as 
\be
\frac{d n_5}{d t}= - n_5 \frac{(2 N_f)^2}{\chi_5}  \frac{\Gamma_{\rm sph}}{2 T} \ ,
\ee
where $\Gamma_{\rm sph}$ is the so-called sphaleron rate.

The sphaleron rate has been computed in Ref.~\cite{Moore:2010jd} at very large $T$ for a $SU(N_c)$ theory, assuming not only a small value of
the QCD coupling constant, but that the inverse of its logarithm should be small. This means that the results of Ref.~\cite{Moore:2010jd} are only strictly valid at very large
values of $T$. The authors of 
Ref.~\cite{Moore:2010jd} suggested that if one could extrapolate their results to the 
 regime where $\alpha_s \sim 1/3$ ---a value more in accordance with what one expects to find in HICs--- one could take
 $\Gamma_{\rm sph} \sim 30 \alpha^4_s T^4$. 
It is worth noticing here that the sphaleron rate has also been computed in a strong coupling regime using AdS/CFT  techniques in Ref.~\cite{Son:2002sd},
finding $\Gamma_{\rm sph}= (g^2 N_c)^4 T^4/256 \pi^3$.
If we assume that the sphaleron rate is correctly described by  $\Gamma_{\rm sph} \sim 30 \alpha^4_s T^4$, 
assuming $\alpha_s =1/3$ one then concludes that the QCD sphalerons start to have an effect at time scales 
 $t_{\rm sph} \sim  0.45/T$. Thus, even in the case that a chiral imbalance is created in the very initial stages
 of the HIC, as suggested in Ref.~\cite{Fukushima:2010vw}, one may expect that this evolves in time in the
 regime where color dynamics is relevant.

The color charges and currents generated in the early moments of the HIC  are rapidly bleached and have a shorter
lifetime than the electromagnetic currents, so we expect they might be absent in the long time behavior of the system. 
In particular, in a weak coupling scenario the color currents are known to disappear at time scales of order $t \sim \frac{1}{\alpha_s \log{1/\alpha_s} T}$  \cite{Manuel:2004gk}, while for moderate to strong coupling
one might expect bleaching to occur in shorter time scales. 
Thus, at longer times one can study the evolution of the quark chirality ignoring the effects of the the color gauge fields, and
taking into account only the electromagnetic fields.  We will assume in the following that a chiral imbalance persists in
 the regime where there are no in-medium chromomagnetic fields.   The axial charge  created has been estimated in models where color flux tubes are generated in the early moments of
HIC~\cite{Kharzeev:2001ev,Lappi:2006fp}, and in the regime where color dynamics can be ignored, values of 
$\mu_5 \sim 100$ MeV can still be considered~\cite{Hirono:2014oda}.

We analyze here the evolution of the chiral magnetic effect in a very simple toy model that would correspond to the time scales where
the color dynamics might be ignored. Our primary goal is to evaluate whether the generic effects we have discussed so far ---which arise in any
electromagnetic conductor where its charged chiral fermions with chiral imbalance are coupled to the electromagnetic fields---
might have any relevance in HICs, rather than presenting a much more realistic evolution of the chiral imbalance in HICs.
Let us mention here that only Ref.~\cite{Tuchin:2014iua} has considered the time evolution of the electromagnetic fields in the QGP in the
presence of a chiral imbalance, however disregarding the  chiral anomaly equation in studying the evolution of the magnetic fields.

\subsection{Toy model and numerical results}
\label{QGPtoy}

In this subsection we study the evolution of the quark chiral imbalance and CME in a very simple toy model. We consider
the presence of very strong magnetic fields, of the order of those found in noncentral HICs, with  $e |{\bf B}| \sim m^2_\pi$~\cite{Skokov:2009qp}.
We will ignore the effects of the gluon fields as these have a shorter lifetime than those of the electromagnetic fields
and currents. In particular, this means that we will ignore the sphaleron rate that allows us to change the quark
chirality. We consider a fixed spherical volume $V$ of radius $R=10$ fm,  made of an isotropic QGP. 
This choice of geometry might look too simple to describe the fireball resulting in a peripheral collision giving rise
to large magnetic fields in the HIC, while a cylinder might look more appropriate. However, our analysis 
 shows that the effect we study does not depend on the geometry of the system. Although our toy model is of fixed volume and assumed to be infinitely long lived, 
our only aim is to see whether the transfer of helicity from the fermions to the gauge fields
can take
place in time scales much shorter than the lifetime of the HIC fireball, as otherwise, it would be irrelevant
for the realistic HIC. We also aim the order of magnitude of the effect.
We should warn the reader that a much more careful analysis valid for HICs should consider
the volume expansion and a more precise spectrum of the magnetic fields which are
obtained in noncentral HICs, which can be obtained with a study similar to  the one carried out in
Ref.~\cite{Gursoy:2014aka}.

For the QGP the chiral density is expressed as 
\be \label{eq:n5mu5} n_5 = N_c N_f \frac{\mu_5 }{24\pi^2} \left( \mu_5^2  + 12 \mu_V^2 +4\pi^2 T^2 \right) \ , \ee
where $N_c=3$ and $N_f=3$ stand for the number of colors and the number of light quark flavors, respectively. 
For simplicity, we assume that $\mu_V=(\mu_R+\mu_L)/2 \ll T$ in all of our computations. Further, for three
light quark flavors, $C= N_c( \frac 49 + \frac19 + \frac19)=2$.

We assume values of the chiral chemical potential of the order $\mu_5 \sim 100$ MeV, as suggested in Ref.~\cite{Hirono:2014oda}, although different
from that reference, we assume that it is homogeneous over the whole volume.
We take values of the electromagnetic conductivity for QCD with three light quark flavors as found in the
lattice QCD $\sigma = 0.0244 \,T$~\cite{Ding:2010ga}, and we will assume temperatures as those found in
the Relativistic Heavy-Ion Collider (RHIC) experiments, $T = 225$ MeV. This means that $\sigma \simeq 5.4$ MeV.

We assume a Gaussian profile in Fourier space for the initial magnetic field, similar to the Gaussian configurations
taken in Refs.~\cite{Hirono:2014oda,Hongo:2013cqa} to do studies of anomalous hydrodynamics for the HIC. 
The precise form of the magnetic spectrum is taken as in Eq.~(\ref{eq:BGaussian}), similar to the one taken in the previous section as a regulator of 
the Dirac delta:
\be \label{eq:gaussfield} B_{{\bf k},0} = b_0 \exp \left[ - \frac{1}{2} \left( \frac{k-k_0}{\kappa} \right)^2 \right] \ . \ee
In our toy model  the peak of the Gaussian is taken to be in the range
between 100--300 MeV, with a width of the order 40 MeV (consistent with the widths in configuration space of Refs.~\cite{Hirono:2014oda,Hongo:2013cqa}). 
This means that if Fourier transformed to configuration space,
the magnetic field is concentrated in a small finite region contained in the volume $V$. 
With this form of the initial spectrum we also guarantee that the typical wave numbers associated to the magnetic field are  bigger than the inverse of the length scale of the size of
the system  $\sim 1/R$.  As in this case we are not interested in taking the infinitely narrow Gaussian, we simply take the amplitude of the magnetic field in such a way that
in configuration space it takes the value
$|e {\bf B}_0|= m_\pi^2$.
Note that  the allowed wave numbers of any magnetic field inside the volume have to be $k > 1/R \sim 20$ MeV, much larger than the value of the electromagnetic
conductivity, $k \gg \sigma$.

We solve numerically Eqs.~(\ref{AMW-1}) and (\ref{AMW-2}) as described in Appendix~\ref{app:numerics}. Let us stress that several scattering processes between quarks and photons, quarks and gluons, and quarks and quarks give a contribution to
the helicity-flipping rate $\Gamma_f$. However they scale as $\alpha^2 m^2_q/T$ and  $\alpha^2_s m^2_q/T$, respectively, and
it can be shown that they are relevant at time scales much larger than the expected lifetime of a HIC fireball $\sim$ 10 fm
(see Appendix~\ref{app:fliprate} for a much more detailed discussion). Because the relevant time scales associated to
the change of the quark chirality due to the chiral anomaly are much shorter, it is reasonable to approximate  $\Gamma_f \approx 0$.

Let us first mention that the occurrence of a chiral magnetic instability is very unlikely, as this requires values of
the magnetic field wavelengths $ k < \frac{C \alpha \mu_5}{\pi}$. If we take the minimal value $k \sim 20$ MeV, then the
initial value of the chiral magnetic potential should be $\mu_{5,0} \geq 4304$ MeV, an extremely large value very unlikely to occur.
For this reason, we ignore this scenario.

 However, a temporary creation of magnetic helicity is indeed possible in a HIC context if an initial chiral imbalance is present in the system. 
We assume a Gaussian shape for the initial magnetic field, centered at $k_0=100$ MeV and width of
$\kappa=40$ MeV. We plot the results of both the axial chemical potential and magnetic helicity in Fig.~\ref{fig:N1_k100_s40_m100_B1_h0} up to large times,
although relevant time scales for the HIC should be $t < 10$ fm.

\begin{figure}[ht]
 \begin{center}
 \includegraphics[scale=0.35]{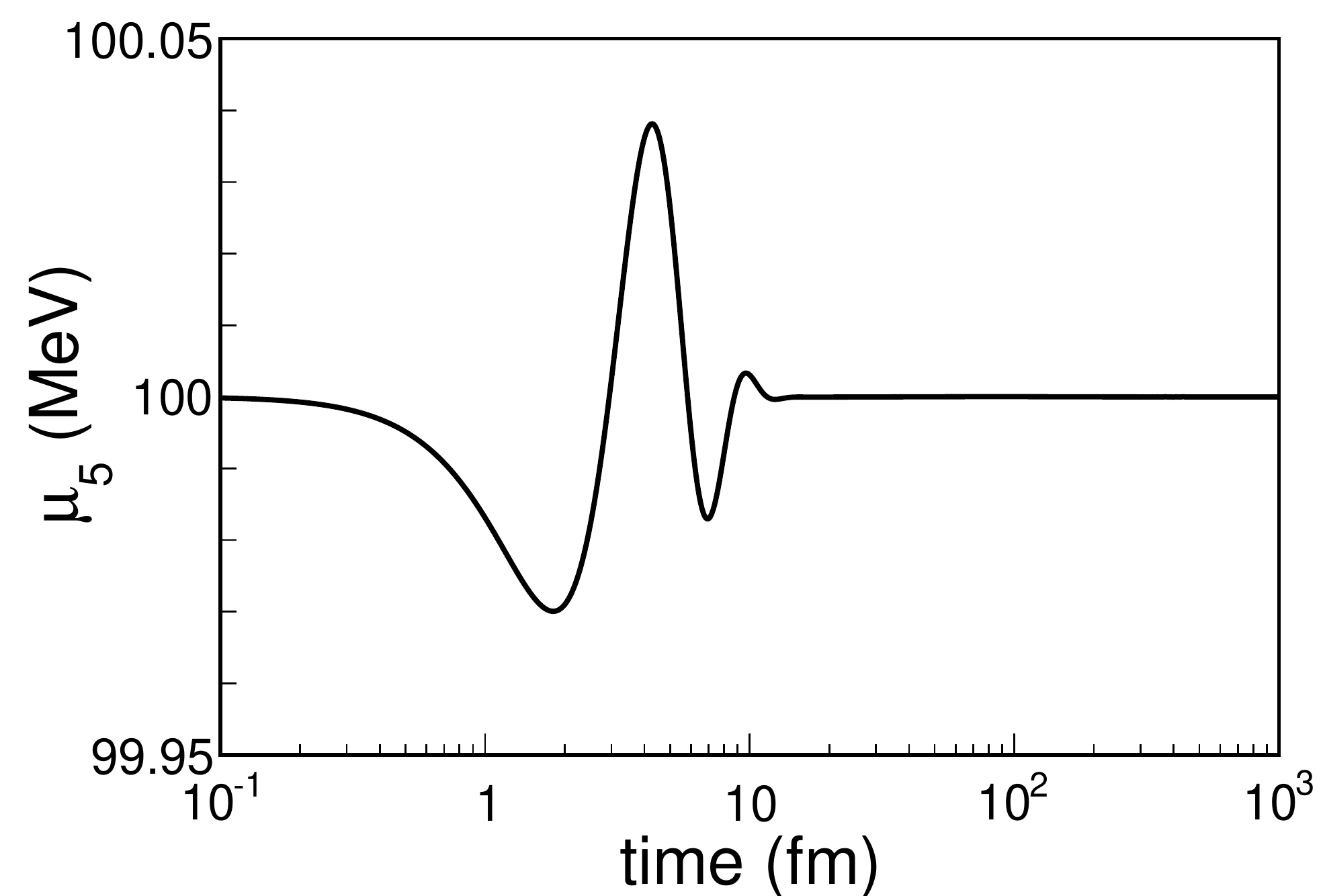}
 \includegraphics[scale=0.35]{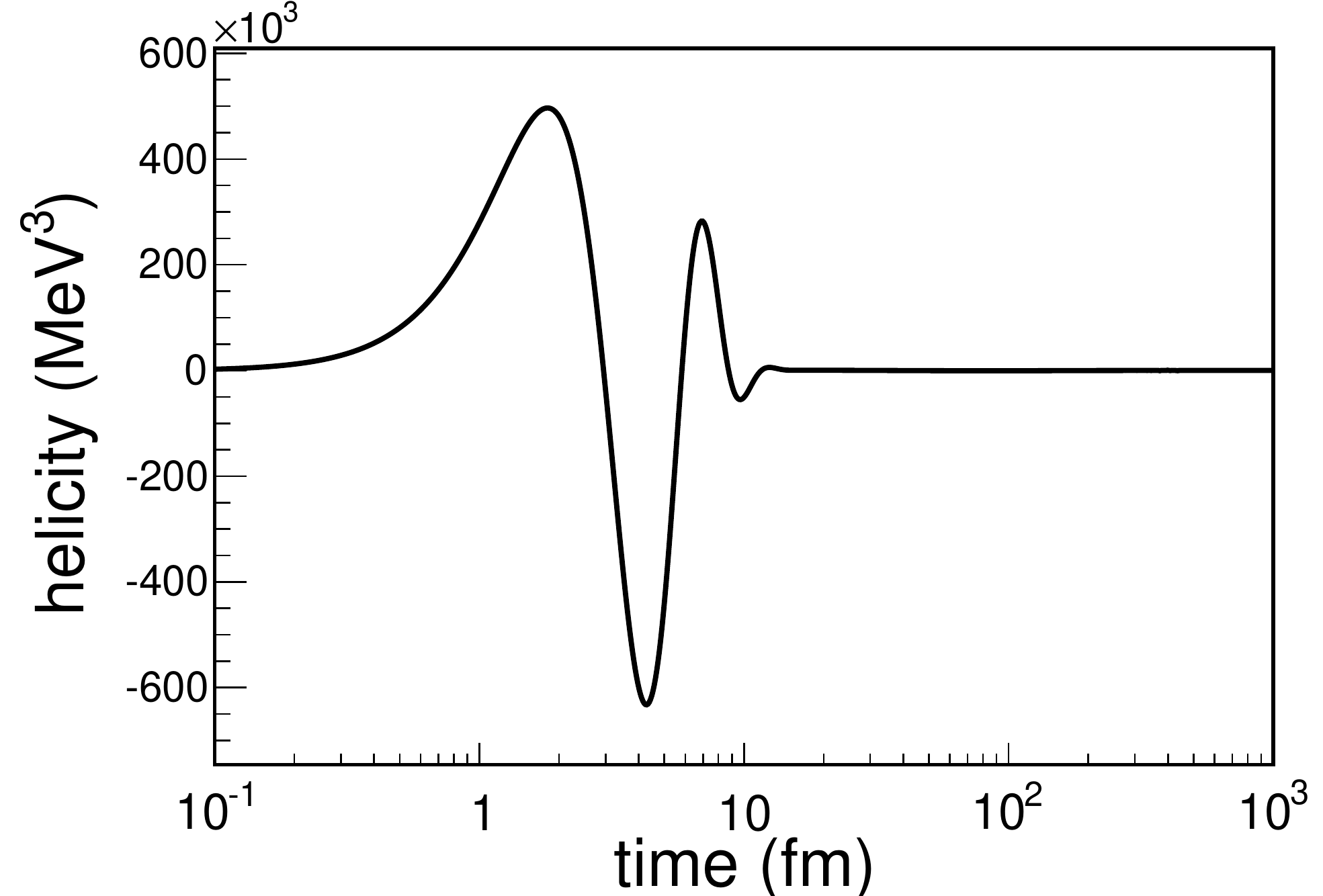}
 \end{center} 
 \caption{\label{fig:N1_k100_s40_m100_B1_h0} Generation of magnetic helicity due to an initial chiral
imbalance. The parameters are $|e {\bf B}_0|=m_\pi^2$, $\mu_{5,0}=100$ MeV and a Gaussian initial magnetic field
centered at $k_0=100$ MeV and width $\kappa=40$ MeV.}
 \end{figure}

We observe the oscillatory behavior of the chiral chemical potential with very small amplitude, of the order of
0.05 \% with respect to the initial value, and the appearance of also an oscillatory magnetic
helicity. These oscillations are characteristic of the solutions for large wave numbers discussed in Sec.~\ref{sec:largek}, although
here we have a continuous spectrum of modes.


Just to check the dependence of the results on the value of the magnetic field, we assume now the same situation with 
a value of the magnetic field as  $|e {\bf B}_0|=10 m_\pi^2$ \footnote{For such large values of the magnetic field one should not obviate the presence of Landau levels, that may modify the electromagnetic current
we have considered so far; we will not do it here, as we only aim to check the dependence of the result on the strength of the magnetic field.}, but using the same parameters of the Gaussian as before and
with the same  chiral imbalance of $\mu_{5,0}=100$ MeV. 
We show our results in Fig.~\ref{fig:N1_k100_s40_m100_B10_h0}. The qualitative shape of the curves does not change. In particular, the periodicity of the oscillation remains invariant
because it is solely controlled by the typical wave number, not by the strength of the magnetic field.  In this case, the axial chemical potential presents larger fluctuations due to the bigger value of the magnetic field. 
We find variations on $\mu_5$ around 4 \% with respect to the initial chiral imbalance, that is, the fluctuations are 100 times larger than the previous example. The magnetic helicity density is also amplified by a factor 
of 100 with respect to the previous example (due to the fact that the magnetic field is chosen to be ten times bigger than before).
\begin{figure}[ht]
 \begin{center}
 \includegraphics[scale=0.35]{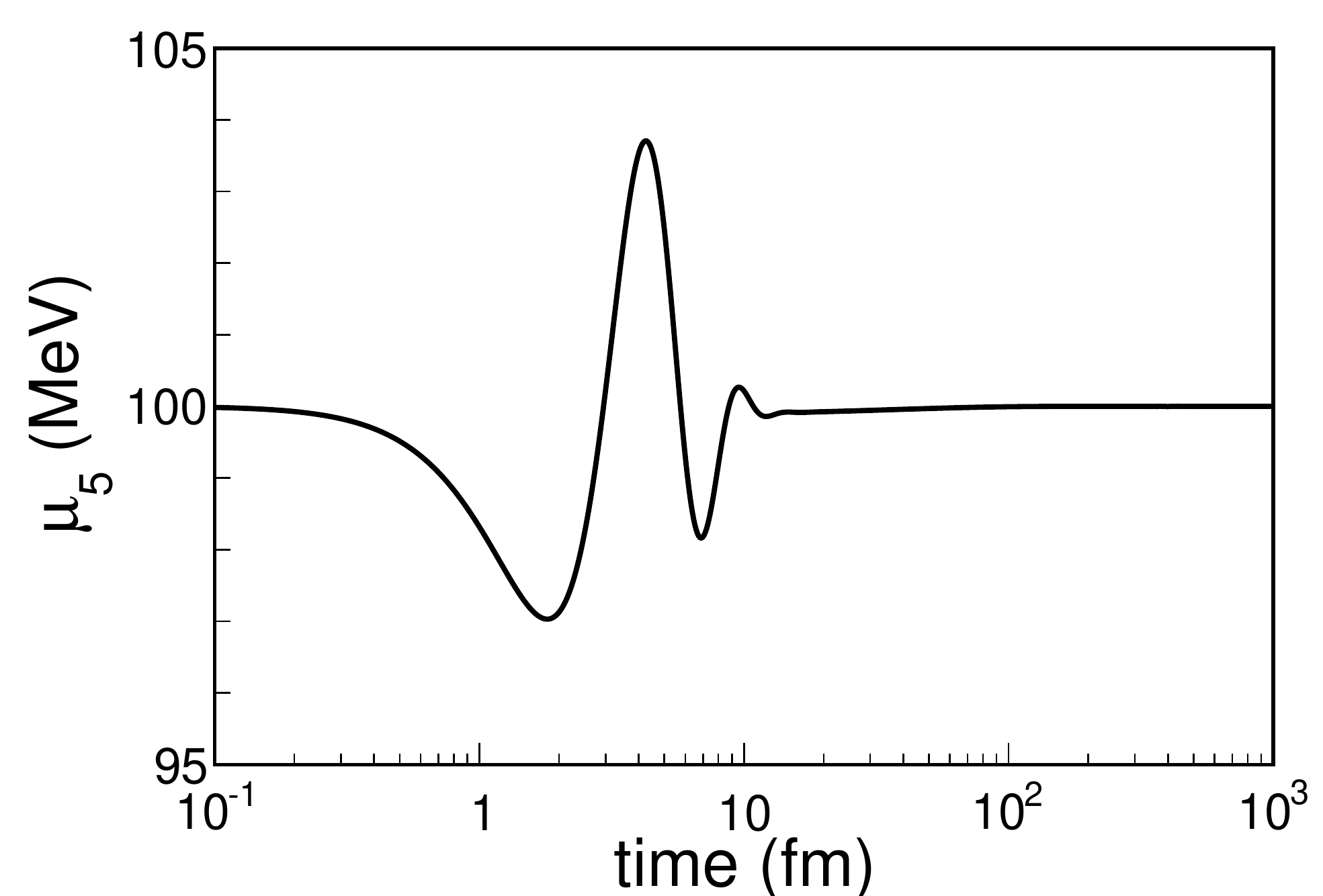}
 \includegraphics[scale=0.35]{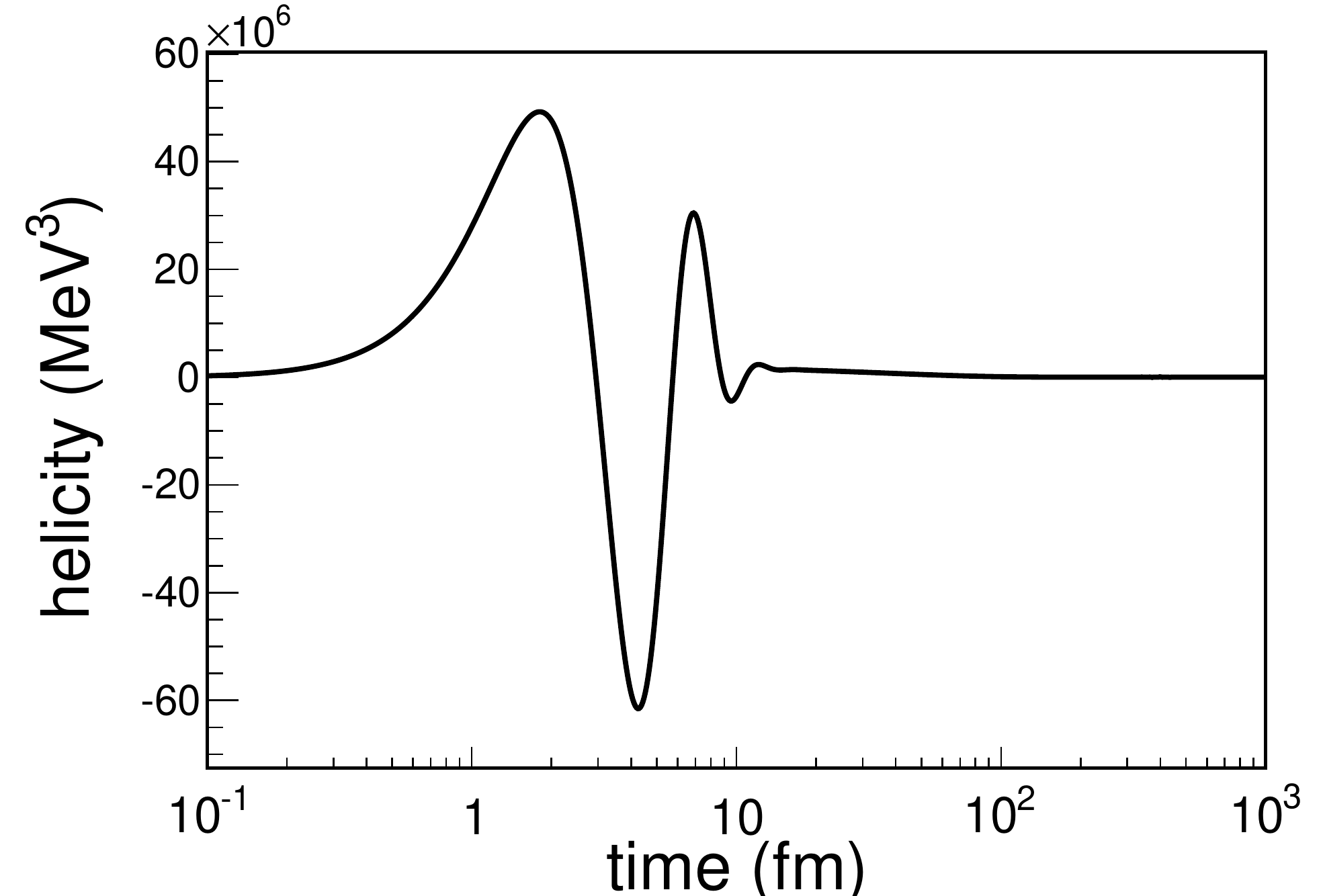}
 \end{center} 
 \caption{\label{fig:N1_k100_s40_m100_B10_h0} Generation of magnetic helicity due to an initial chiral
imbalance. The parameters are $|e {\bf B}_0|=10m_\pi^2$, $\mu_{5,0}=100$ MeV and a Gaussian initial magnetic field
centered at $k_0=100$ MeV and width $\kappa=40$ MeV.}
 \end{figure}

We now test the dependence of our results to the value of $k_0$. With initial $\mu_{5.0}= 100$ MeV and magnetic field $|e {\bf B_0}|=m_\pi^2$, we take $k_0=200$ MeV. The results are shown in
Fig.~\ref{fig:N1_k200_s40_m100_B1_h0}. One can see an increase of the oscillation frequency with respect to Fig.~\ref{fig:N1_k100_s40_m100_B1_h0}.
This is because when increasing the typical wave number $k_0$, the factor $\Delta_\pm$ grows accordingly, and the
sinusoidal oscillations have smaller periodicity.
\begin{figure}[ht]
\begin{center}
\includegraphics[scale=0.35]{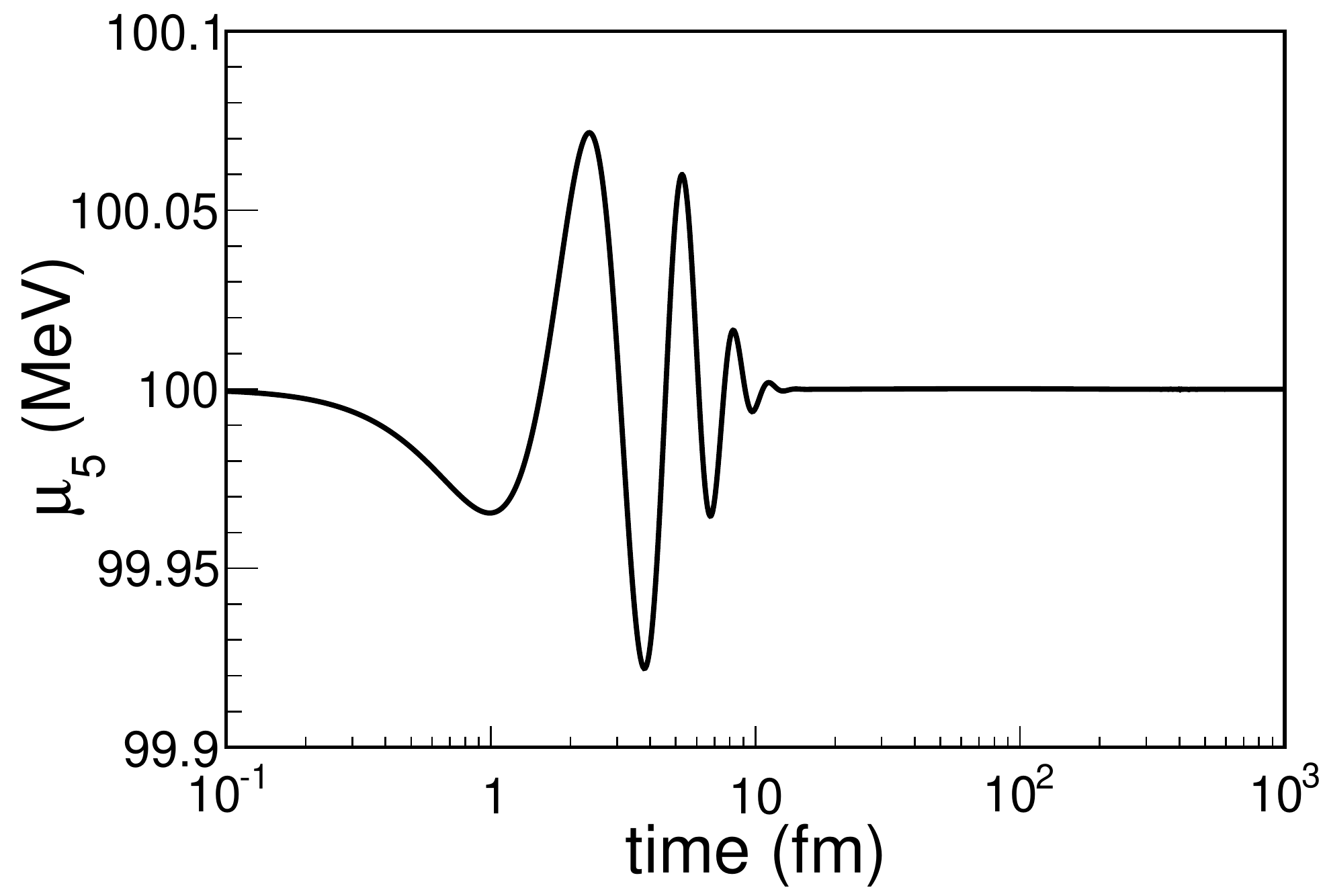}
\includegraphics[scale=0.35]{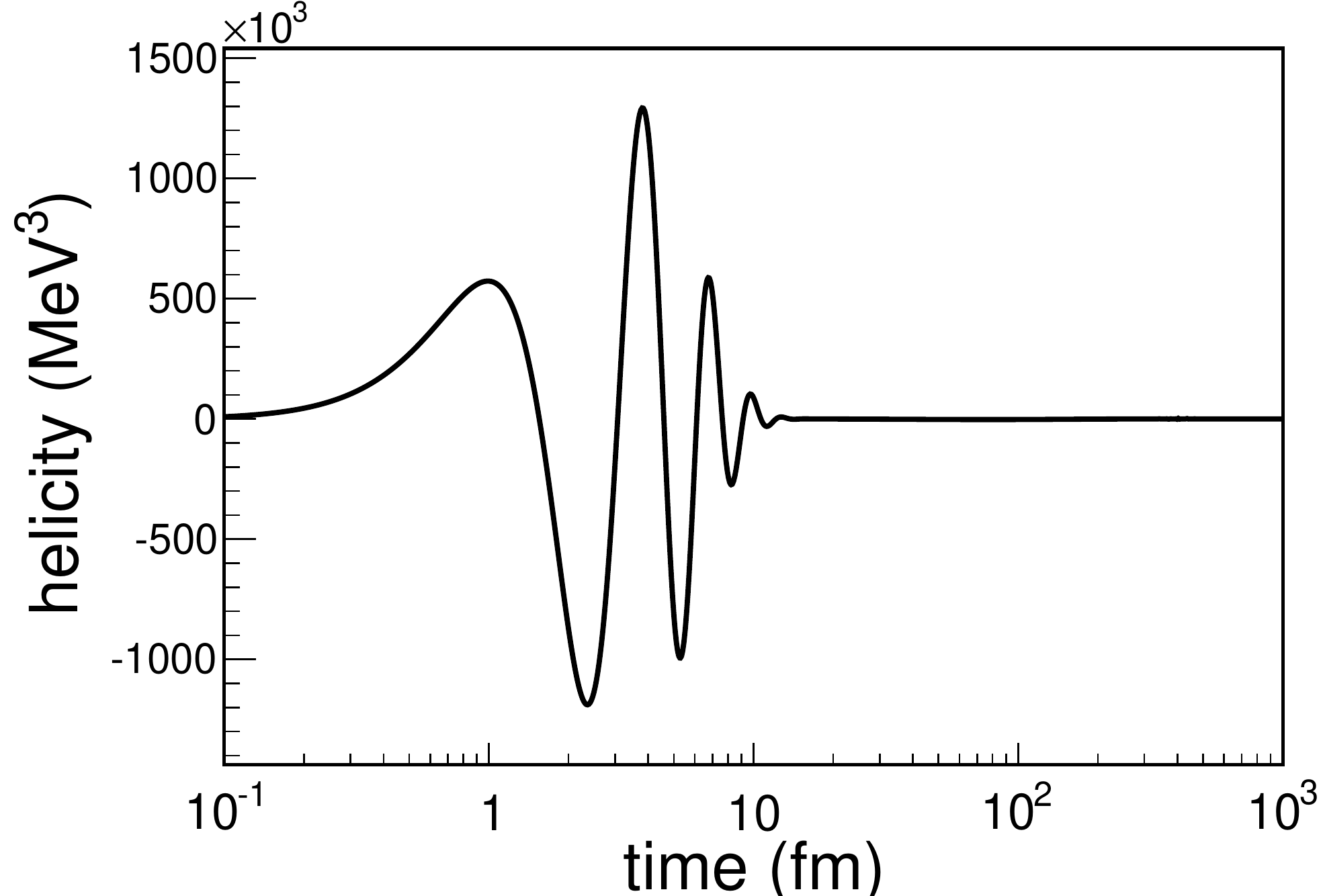}
\end{center} 
\caption{\label{fig:N1_k200_s40_m100_B1_h0} Generation of magnetic helicity due to an initial chiral imbalance. The parameters
are $|e {\bf B}_0|=m_\pi^2$, $\mu_{5,0}=100$ MeV and an initial magnetic field taken to be Gaussian
centered at $k_0=200$ MeV and width $\kappa=40$ MeV.}
\end{figure}

To conclude the study of our anomalous Maxwell equations we present the case with a dynamical generation of chiral
imbalance due to an initial magnetic helicity in the system. In this case we impose $\mu_{5,0}=0$ MeV and a left-handed polarized
magnetic field consistent with $|e {\bf B}_0|=m_\pi^2$, being the right-handed polarized component kept to zero in order to force
a nonzero initial magnetic helicity. Our results are presented in Fig.~\ref{fig:N1_k100_s40_m0_B1_h1}.
\begin{figure}[ht]
 \begin{center}
 \includegraphics[scale=0.35]{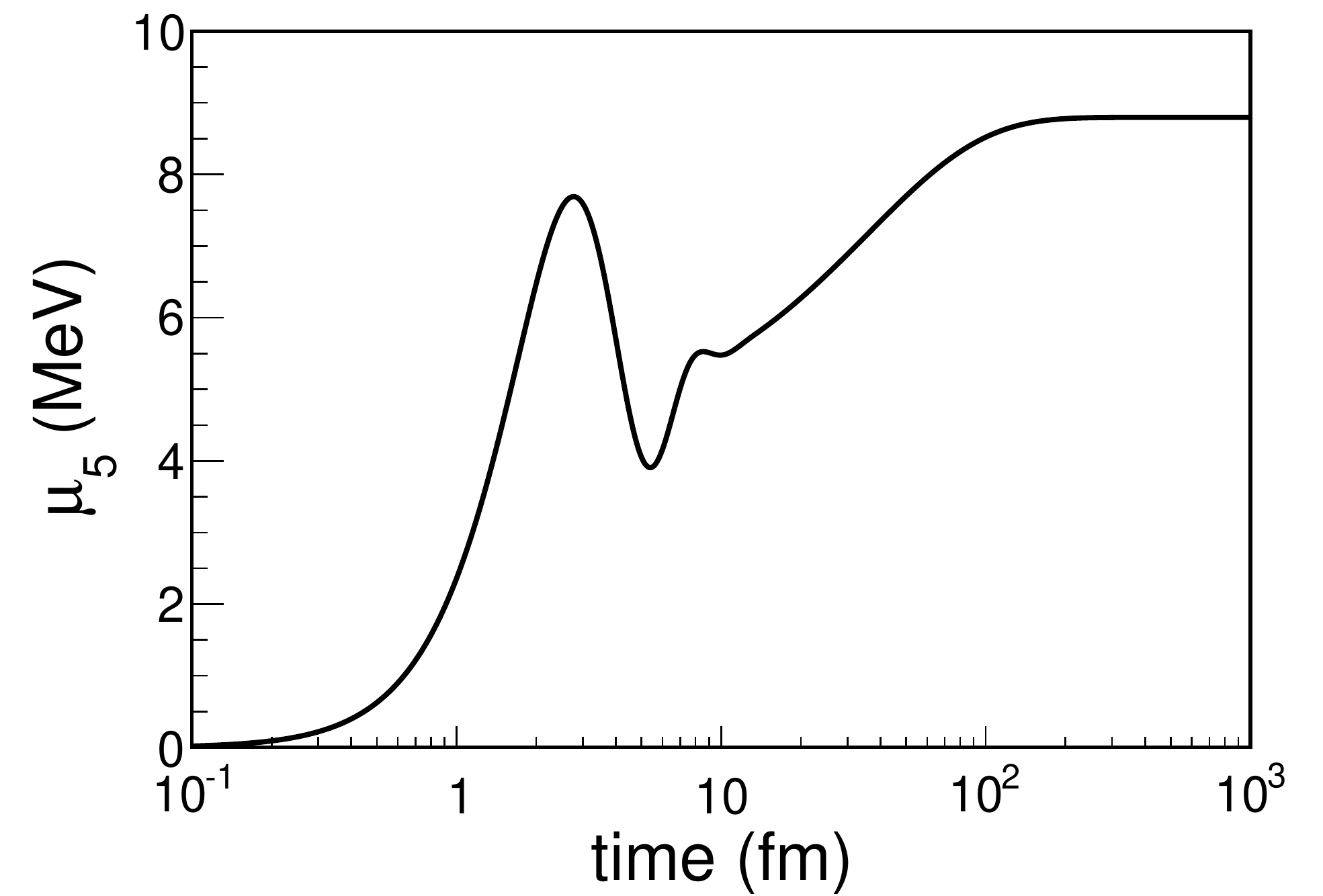}
 \includegraphics[scale=0.35]{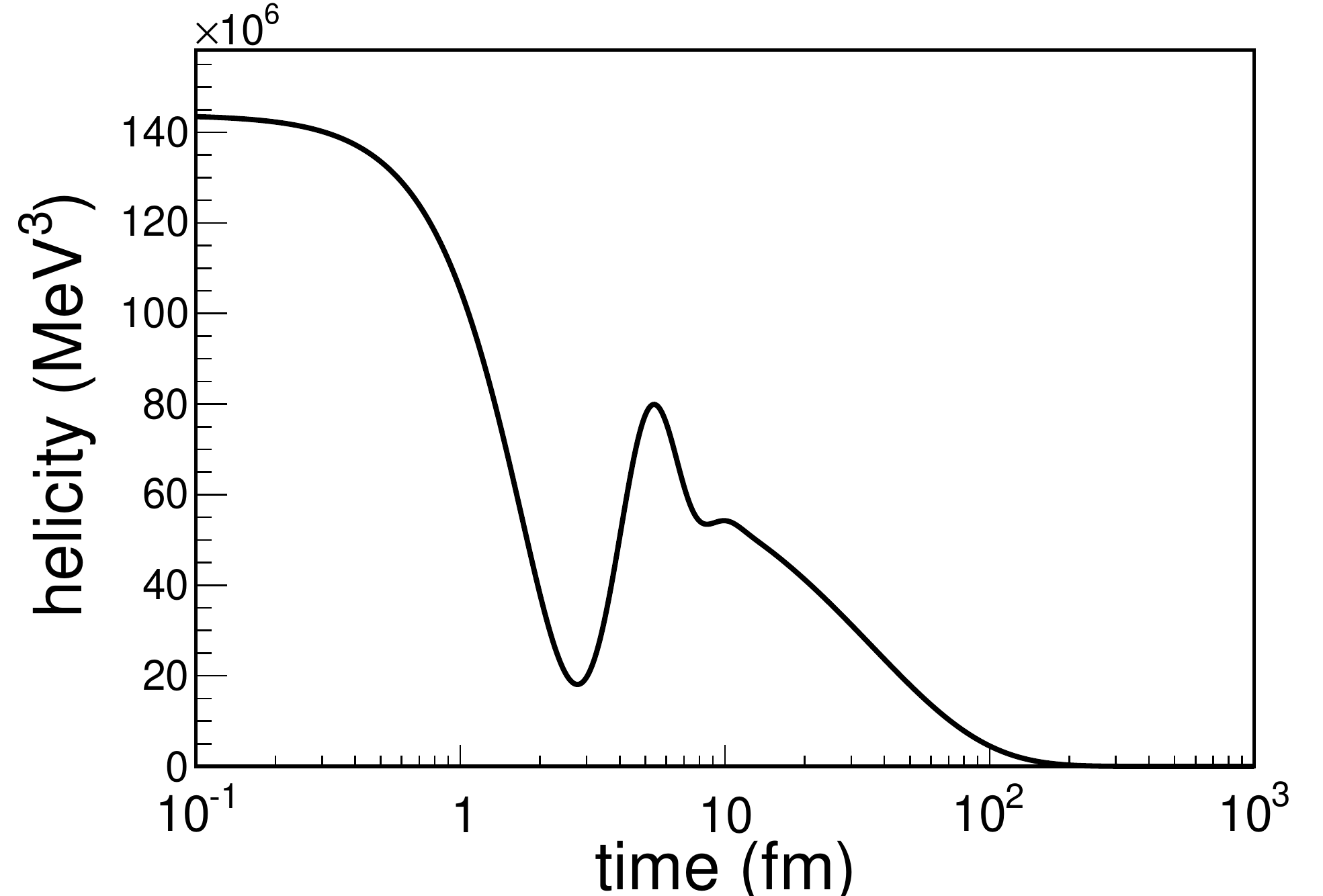}
 \end{center} 
 \caption{\label{fig:N1_k100_s40_m0_B1_h1} Generation of chirality imbalance due to an initial magnetic helicity.
The parameters are: $|e {\bf B}_0|=m_\pi^2$, $\mu_{5,0}=0$ MeV and an initial magnetic field taken to be Gaussian
centered at $k_0=100$ MeV and width $\kappa=40$ MeV.}
 \end{figure}

We observe a generation of chiral imbalance at the cost of the decrease of the initial magnetic
helicity. The effect continues until times as large as 100 fm (or possibly earlier, depending on the presence of some
nonzero $\Gamma_f$). Finally we study the same generation of chiral imbalance when the Gaussian profile is centered at a higher wave number, $k_0=200$ MeV. 
We show the numerical results in Fig.~\ref{fig:N1_k200_s40_m0_B1_h1}. 
\begin{figure}[ht]
 \begin{center}
 \includegraphics[scale=0.35]{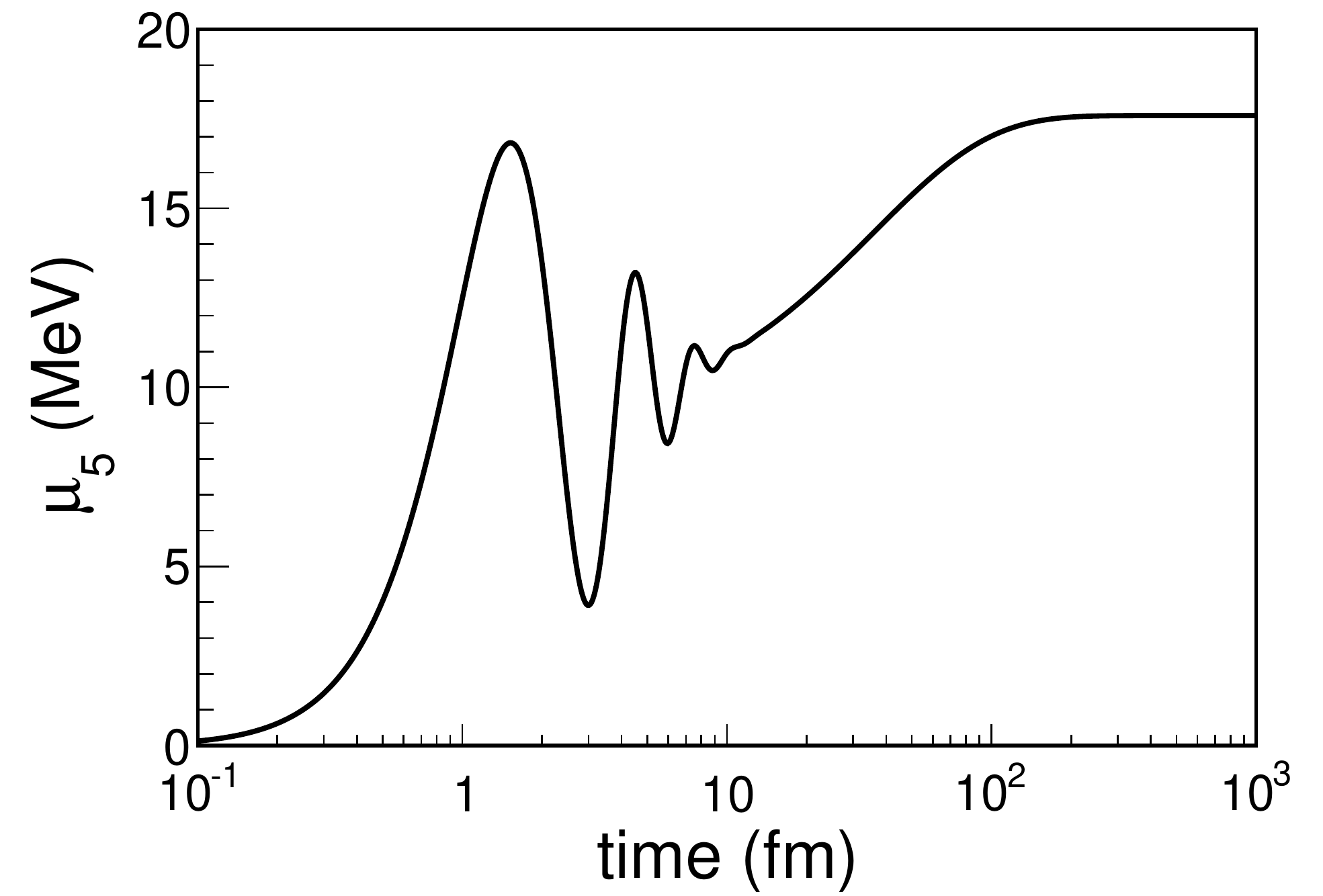}
 \includegraphics[scale=0.35]{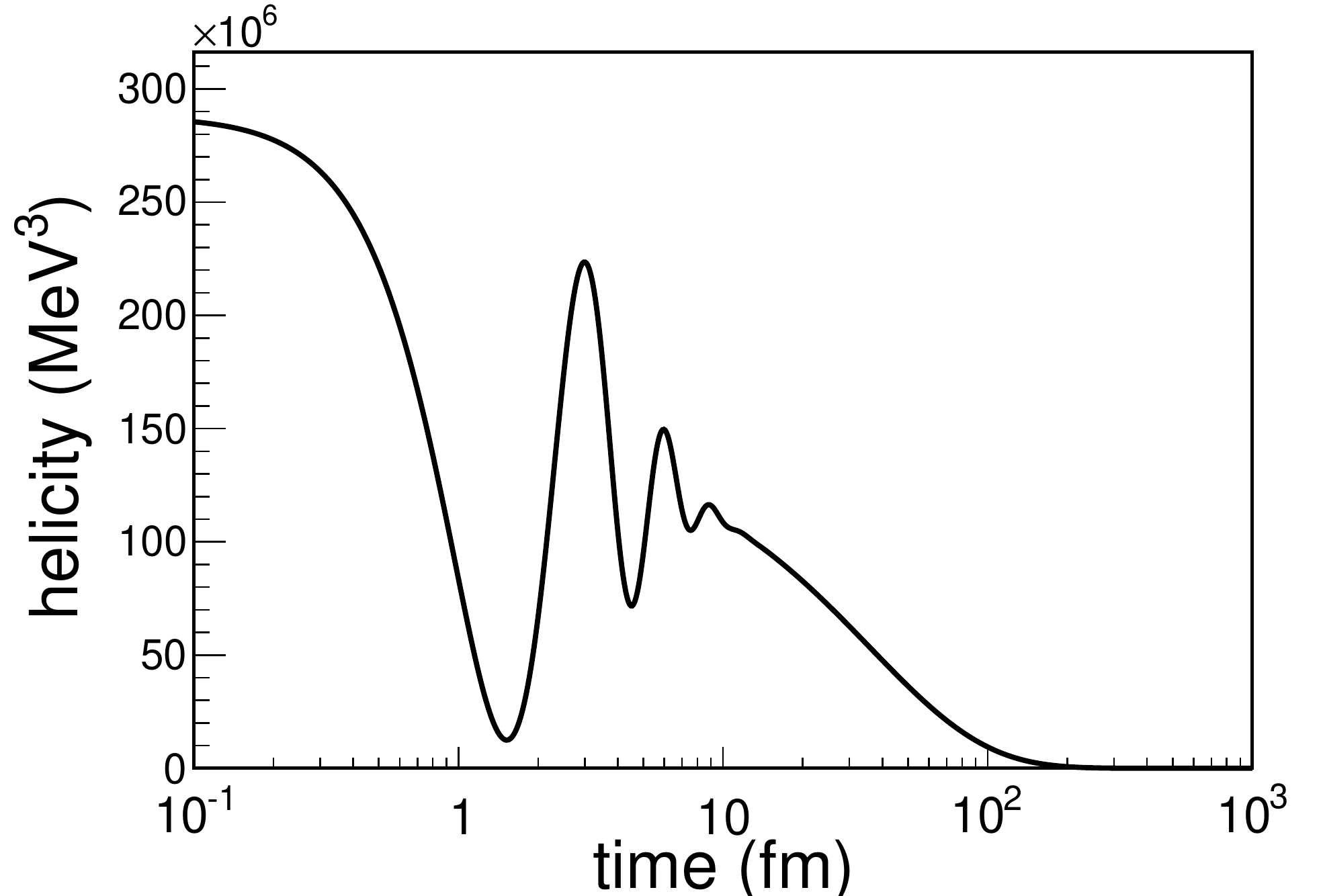}
 \end{center} 
 \caption{\label{fig:N1_k200_s40_m0_B1_h1} Generation of chirality imbalance due to an initial magnetic helicity.
The parameters are $|e {\bf B}_0|=m_\pi^2$, $\mu_{5,0}=0$ MeV and an initial magnetic field taken to be Gaussian
centered at $k_0=200$ MeV and width $\kappa=40$ MeV.}
 \end{figure}

From the past examples, we expect an increase of the frequency of the oscillations. In addition, we observe an increase of the generated chiral imbalance. This amplification is 
affected from the increase (a factor of 2) of the initial magnetic helicity. 
To understand this factor we can compute the initial magnetic helicity density substituting Eq.~(\ref{eq:gaussfield}) for the left-polarized magnetic field into Eq.~(\ref{eq:defhelicity}) (and setting the
right-polarized component to zero). Performing the integration and using Eq.~(\ref{eq:b0ampl}) we obtain
\be {\cal H}_0 \simeq \frac{|e {\bf B}_0|^2  \pi^{1/2}}{\alpha \kappa^5 V} k_0 \ , \ee
where the relation is valid for $k_0$ moderately larger than $\kappa$. Therefore, the initial magnetic helicity
is roughly proportional to $k_0$, which accounts for the amplification factor of the magnetic helicity density in the last example.
As opposed to the results in Sec.~\ref{sec:smallk} (where we neglected the second time derivative in the AME), the tracking chiral
imbalance does not depend explicitly on $k_0$, only indirectly through the initial magnetic helicity density. 
Therefore, we want to remark that a more precise knowledge of the
initial magnetic field profile is needed to determine a solid conclusion of this effect in HIC experiments.
Note also that we have only shown the time evolution of the magnetic helicity, and not of the single magnetic field
components, which might evolve differently depending on its initial geometrical configuration. A much more realistic
configuration for the HIC would have implied to choose the initial magnetic field created by moving charges in one direction, as done, for example,  
in Ref.~\cite{Tuchin:2014iua}. However, as the amount of helicity is governed by the value of $\mu_5$ and the chiral anomaly equation,
we expect that our plots for ${\cal H}$ would not change drastically.

\section{Conclusions}
\label{conclu}
     
  In this work we have studied the coupled system of Maxwell and chiral anomaly equations valid for EM conductors made up by chiral fermions.
We have proven that a possible derivation of this coupled set of equations can be performed in the framework of chiral kinetic theory in the relaxation time approximation.
The resulting set of anomalous equations are fully general, and can have a wide range of applications in systems so different such as  electroweak and  quark-gluon plasmas
or in condensed matter systems such as Dirac and Weyl semimetals.

We have considered the AME to study the evolution of both magnetic field and fermion chiral imbalance.
After integrating the chiral anomaly equation over space on a finite volume,  a conservation law relating the chiral charge and  magnetic helicity densities is 
obtained. When expressed in Fourier space and in circular polarized components, we observe that the dynamical evolution of the left- and right-handed components of
the magnetic field in the presence of a finite $\mu_5$ is different. Then, it is easy to understand that in the presence of a fermionic chiral imbalance  a nonvanishing  value of the magnetic helicity, which measures
the asymmetry between the two different polarizations, soon arises. We have carried out a detailed study of the transfer of helicity from fermions to magnetic fields, with different initial conditions, and realized that the rate of this
transfer depends drastically on the characteristic wave number of the magnetic field in consideration, while the amount depends on the magnitude of the initial magnetic field.

Using the same set of equations but with different initial conditions, we also showed that a chiral equilibrated plasma might develop an imbalance between right- and left-handed fermion populations
if a nonzero magnetic helicity is applied to the system. This occurs whenever there is a mechanism that damps the magnetic helicity, in the case we are studying,
the Ohmic dissipation. Then the chiral anomaly equation dictates that the lowering of the magnetic helicity comes together with an increase in the chiral fermion imbalance.
 The equations tell us that the creation of fermion imbalance persists in the system as long as no other dynamical process erases it.
We examined this case for systems with characteristic small and large frequencies, and found the rate of the transfers in each case.

We have focused our attention to the QGP that is supposed to occur in heavy-ion collisions. While it has been recognized
that the CME would have a clear impact on the charge-dependent hadron azimuthal correlations detected both in the 
RHIC and in the LHC for noncentral collisions, our main claim is that the CME changes the dynamical evolution of the magnetic field in a way that might affect the studies of those correlations performed
to date~\cite{Hirono:2014oda,Hongo:2013cqa}. A similar conclusion was reached in Ref.~\cite{Tuchin:2014iua}, after studying the magnetic field originated by a point particle in the QGP in the 
presence of the CME current, where also rapid oscillations of the magnetic field were obtained, even if the time evolution of $\mu_5$ was ignored.

While we realize that the occurrence of a chiral magnetic instability is very unlikely to occur in HICs~\cite{Akamatsu:2013pjd} ---as it would require extremely large values of the chiral
chemical potential--- a finite nonvanishing magnetic helicity would soon arise if a fermion chiral imbalance persists in the late stage evolution of the QGP. The magnetic helicity gives account
of a rather nontrivial topology of the magnetic field lines, with formation of knots and helical structures. While our conclusions have been extracted from a simple toy model, a much more careful
analysis should be carried out for applications to HICs.
Then, another relevant consequence of the the chiral fermion imbalance is that it would affect the estimate of the lifetime of the magnetic fields in a HIC. This is so because
with magnetic helicity a magnetic field configuration tends to be more long lived~\cite{Candelaresi:2011pg}.

The possibility of a chiral imbalance generation due to an initial presence of helical fields also deserves more attention in the context of HICs. This case requires a deeper knowledge of the topological arrangement of the initial magnetic 
fields created by spectator nucleons. Any difference on the right- and left-handed polarized components of the magnetic field might potentially cause nontrivial structures as knots, that induce the creation of a quark chiral asymmetry.
It is important to recognize that this generation of chiral imbalance has an Abelian origin and could represent an alternative scenario
to the one proposed so far, which is intimately linked to the topological properties of a non-Abelian gauge theory such as QCD.

We leave for future work a more realistic study of the anomalous magnetohydrodynamic equations in the
context of HICs. For this goal, one can follow the lines of Refs.~\cite{Voronyuk:2011jd, Konchakovski:2012yg,Toneev:2011aa, Toneev:2012zx},
where the particles of the HIC are described by a dynamical quasiparticle model.
The electromagnetic fields are computed from the Li\'enard--Wiechert potentials for all charged quasiparticles (both
spectators and participants). Likewise, these retarded EM fields produce a backreaction on the quasiparticles which
are propagated in time. In this way, the spacetime evolution of electromagnetic fields can be computed with a
realistic expansion of the fireball. The extension of this model to include a chiral imbalance will be
addressed in the future.

\begin{acknowledgments}
We acknowledge discussions with A.A.~Andrianov and D.~Espriu.
 This research was in part supported from Ministerio de
Ciencia e Innovaci\'on under contracts No. FPA2010-16963 and No. FPA2013-43425-P, and Programme TOGETHER from R\'egion Pays de la Loire and the European I3-Hadron Physics program.
\end{acknowledgments}

\appendix

\section{Chiral kinetic theory in the relaxation time approximation}
\label{app:secCKT}

In this section we show how the AMEs studied in the remaining part of the manuscript can be
derived from CKT. First, we will briefly  review the basic inputs of the (semi)classical kinetic theory (see 
Refs.~\cite{Son:2012wh,Stephanov:2012ki,Son:2012zy,Chen:2012ca,Manuel:2013zaa,Manuel:2014dza} for a more detailed
description).

The distribution function associated to a chiral fermion $f_p$ obeys the transport equation
 \be
 \label{Boltzmanneq}
  \frac{df_p}{dt} = \frac{\pa f_p}{\pa t} +  \dot{ {\bf r}} \cdot \frac{\pa f_p}{\pa {\bf r}} +  \dot{ \bf{p}} \cdot \frac{\pa f_p}{\pa {\bf p}} = C[f_p] \ ,
 \ee
where $C[f_p]$ is the collision term. The fermion equations of motion take into account the effects of the Berry curvature and 
for a fermion of charge $e$ read
\ba
\label{dr/dt}
 (1 + e  {\bf B}\cdot {\bf \Omega_p}) \dot{\bf r} & = & \tilde {\bf v} + e   ( \tilde{\bf E}  \times {\bf \Omega_p})
 + e  {\bf B} (\tilde{\bf v} \cdot {\bf \Omega_p})  , \\
 \label{dp/dt}
 (1 + e  {\bf B}\cdot {\bf \Omega_p}) \dot{\bf p} & = &  e ( \tilde{\bf E}  + \tilde {\bf v}\times {\bf B}) + e^2  {\bf \Omega_p}  ( \tilde{\bf E}  \cdot {\bf B}) ,
  \ea
where we have defined 
$
 \tilde{\bf E}  \equiv {\bf E}- \frac 1e \frac{\partial \epsilon^+_{\bf p}}{\partial \bf r} $ and
$\tilde {\bf v}  \equiv \frac{\partial \epsilon^+_{\bf p}}{\partial {\bf p}} $ and the Berry curvature is expressed as
\be
\Om = H \frac{{\bf p}}{2 |{\bf p}|^3} \ ,
\ee
where  $H=\pm 1$  gives account of the helicity of the  right- and left-handed fermions, respectively.
The energy  for particles/antiparticles is also defined as \cite{Son:2012zy,Manuel:2014dza}
\be
\label{mod-dl}
\epsilon_p^{\pm} =\pm p (1 - e  \Bvec \cdot  \Om ) \ ,
\ee
respectively.



Then it is possible to show
that the axial current defined as  $j^\mu_A = j^\mu_R - j^\mu_L$
obeys the chiral anomaly equation~\cite{Son:2012wh,Stephanov:2012ki,Son:2012zy,Chen:2012ca,Manuel:2013zaa,Manuel:2014dza}
\be 
\label{chiral-anomaly}
\pa_\mu j^\mu_A =  \frac{e^2}{2\pi^2 }  {\bf E} \cdot {\bf B} 
\ , 
\ee
while the vectorial current $j^\mu_V = j^\mu_R + j^\mu_L$ is conserved.

The system of equations presented here has been analyzed in the collisionless limit using a linear response analysis in
Refs.~\cite{Son:2012zy,Manuel:2013zaa}. There, it has been checked that the resulting EM current reproduces both the nonanomalous
and anomalous hard-thermal-loop/hard-dense-loop effective actions that appear in a quantum field theory approach.

The next natural step is to see the effects of collisions in the dynamics. So far, collision kernels in chiral kinetic theory have not been constructed, although it has been pointed out
that they should be nonlocal in order to respect Lorentz invariance~\cite{Chen:2014cla}. However, in order to see the effects of collisions at large times, one typically
uses the so-called relaxation time approximation (RTA). 
Although the RTA gives a good qualitative behavior of the solutions of the Boltzmann equation at large times, it is well known that it does not respect the conservation laws of the system.
In Ref.~\cite{Bhatnagar:1954zz} an improvement to the RTA collision term was proposed, such that it does not spoil particle number conservation. We propose this alternative form for the relaxation-time collision term in CKT, not to violate vector current conservation, nor the chiral anomaly equation.
Hence, we will use a collision term of the Bhatnagar--Gross--Krook (BGK) type~\cite{Bhatnagar:1954zz}
\be
C^{\rm BGK}[ f_p] =- \frac{1}{\tau} \left[ f_p - \frac{n}{n_{eq}} f^{eq}_p \right] \ .
\ee
It is easy to prove that $ \int \frac{d^3 p}{(2 \pi)^3 }  (1 + e \Bvec \cdot  \Om )C^{\rm BGK}[ f_p] =0$, and thus that particle number conservation is not affected by the presence of the
collision term. We note here that in Ref.~\cite{Satow:2014lva} a different form of the collision term was used. The authors checked {\it a posteriori} that the solution they found
was respectful with the particle conservation laws. We note that the proposal we present above gives the same final answers, but in a much faster way.

To study the effects of the collisions in the dynamics we proceed as in Refs.~\cite{Son:2012zy,Manuel:2013zaa}.
The distribution function is expanded in powers of $e$ up to linear order 
\be f_p = f^{eq}_p + e \left( \delta f_p^{(\epsilon)} + \delta f_p^{(\epsilon \delta)} \right) \ , \ee
and the transport equation is solved in a perturbative expansion, assuming that the vector gauge fields are of order $A_\mu = {\cal O}(\epsilon)$, while the spatial derivatives
are $\partial/\partial x^\mu = {\cal O}(\delta)$.
Inserting this expansion into the kinetic equation, the relaxation time modifies the expression of the out-of-equilibrium distribution functions.
The solution can be easily computed in Fourier space
\ba
 \delta f_p^{(\epsilon)} &=& - \frac{ {\bf E} \cdot {\bf v}}{i v\cdot k + \tau^{-1} } \frac{df_p^{eq}}{dp} \ - \tau^{-1} \frac{f_p^{eq}}{iv\cdot k + \tau^{-1} } \left( 1 - \frac{n^{(\epsilon)} }{n_{eq}} \right) \ , \\
 \delta f_p^{(\epsilon \delta)} &=& H \frac{ {\bf v} \cdot \pa_t {\bf B}}{i v \cdot k + \tau^{-1} } \frac{1}{2p} \ \frac{df_p^{eq}}{dp}    \ .
 \ea
where $n^{(\epsilon)}$ is the particle density computed up to order ${\cal O}(\epsilon)$. This solution
takes the same form as expressed in the collisionless limit (see Eqs.~(36) and (38) of Ref.~\cite{Manuel:2013zaa}) 
but with the replacement $i v \cdot k  \rightarrow i v \cdot k + \tau^{-1}$ in the denominators of the expressions, plus one new additional term
%
%
\be J^{(\epsilon) \mu}_{\rm extra}  = - \frac{e^2}{\tau} (n^{eq} - n^{(\epsilon)}) \int \frac{d\Omega_v}{4\pi} \frac{v^\mu}{i v \cdot k + \tau^{-1}} \ , \ee
where $d\Omega_v$ is the solid angle element.

Out of equilibrium, we parametrize the right-/left-handed particle densities in a similar way as in thermal equilibrium,
\be
 n^{(\epsilon)}_{R/L} = \frac{1}{6\pi^2} \left( \mu^{2}_{R/L} + \pi^2 T^2 \right) \mu_{R/L} \ ,
 \ee
where we have introduced out-of-equilibrium chemical potentials.
Defining $\delta \mu = \mu - \mu^{eq}$ we then express 
$ n^{eq} - n^{(\epsilon)} = \left( \frac{T^2}{3} + \frac{\mu^2}{\pi^2} \right) \delta \mu \equiv  \chi \delta \mu$.
 It is then possible to show that the solutions we have found here agree exactly
with those of Ref.~\cite{Satow:2014lva}.

In a self-consistent treatment of the system, the EM fields which appear in the transport equation are generated by the same charged fermions. Thus, they should obey Maxwell equations
\be
\label{Maxwell}
\pa_\mu F^{\mu \nu} = J^\nu \ .
\ee
Assuming that the system is formed by up to $N_s$ different particle species, the total electromagnetic current is expressed as
\be
J^\mu = (\rho, {\bf J}) = \sum_{s= 1}^{N_s} e_s \left[  ( n_s, {\bf j}_s) \,  -   (\bar{ n}_s, \bar {{\bf j}}_s) \right]  \ , 
\ee
where $e_s$ is the charge associated to a given species of particles, and for a relativistic plasma
we have included both fermions and antifermions with different chiralities of different species.

We are interested in the solutions in the limit where $ i v \cdot k \ll \tau^{-1}$. 
Then one finds the total EM current
\be
\label{elcurrentsta}
 {\bf J} =  \sigma  {\bf E} + \sum_{s= 1}^{N_s} \frac{e_s^2 \, \mu_5}{4 \pi^2} \, {\bf B} \ ,
  \ee
 where $\mu_5 = \mu_R -\mu_L$, $\sigma =\tau m_D^2/3$ is the electrical conductivity of the plasma, 
$m^2_D = \sum_s e^2_s \left( \frac {T^2}{3} + \frac{\mu_R^2 + \mu_L^2}{2 \pi^2}\right)$ is the Debye mass, and
we have summed the contribution of every fermion species. Thus, we see that the EM current at large times gives account of both
Ohm's law and the CME.

\section{Numerical study of the AMEs\label{app:numerics}}

 We present here some details on how we have performed the numerical study of the AME equations.
Let us recall that for $k \gg \sigma$ one needs to consider the second time derivative in Eq.~(\ref{AMW-1}). However, in this case an analytic solution analogous to Eq.~(\ref{SolAMW-1}) does not exist for a nonconstant $\mu_5$.
The system of equations (\ref{AMW-1}-\ref{AMW-2}), together with a relation between $n_5$ and $\mu_5$, such as  the relation~(\ref{eq:n5mu5}) for the QGP,  must be solved numerically to obtain $B_{{\bf k}}^+ (t),B_{{\bf k}}^- (t)$ and $\mu_5(t)$. We discretize the time, $t_i,t_i+\Delta t,...,t_f$ and apply 
the following scheme to solve the system:
 
 \begin{enumerate}
  \item We set initial conditions for the chiral imbalance $\mu_5(t_i)$ and the magnetic fields $B_{{\bf k}}^\pm (t_i)$. The latter is taken to have a Gaussian spectrum peaked at $k_0$ 
  \be \label{eq:BGaussian} B_{{\bf k}}^\pm (t_i) = b_0 \exp \left[ - \frac{1}{2} \left( \frac{k-k_0}{\kappa} \right)^2 \right] \ ,  \ee
  where the width of the Gaussian is controlled by $\kappa$.
  
  In this work we have used two different criteria to fix the value of $b_0$:
   \begin{itemize}
    \item In Sec.~\ref{sec:smallk} we solve the AMEs in the low wave number limit using a Dirac delta as an initial condition for the magnetic field squared. 
    To be consistent with this choice in the high wave number case we use
   \be b_0 = \sqrt{\frac{k_0}{\sqrt{\pi} \kappa}} \ |{\bf B}_0 | V \ . \ee
   This value allows us to connect the Gaussian profile with a Dirac delta spectrum in the limit
  \be \lim_{\kappa \rightarrow 0} |B^\pm_{{\bf k}} (t_i)|^2 =  k_0 |{\bf B}_0|^2 V^2  \lim_{\kappa \rightarrow 0} \frac{1}{\sqrt{\pi} \kappa}  \exp \left[ - \left( \frac{k-k_0}{\kappa} \right)^2 \right] = k_0 |{\bf B}_0 V|^2  \delta(k-k_0) \ , \ee
  in consistency with Eq.~(\ref{eq:monocrom}) of Sec.~\ref{sec:smallk}. 
  \item  For the QGP we do not attempt to take the Dirac delta limit. Therefore, we use more realistic (wider) Gaussians instead. For this reason, we just fix $b_0$ in such a way that performing an inverse Fourier transform we recover the value $|{\bf B}_0|$
  in configuration space. Therefore,
  \be \label{eq:b0ampl} b_0 = \frac{|{\bf B}_0| (2\pi)^{3/2}}{\kappa^3} \ \ee
  for all cases in Sec.~\ref{QGPtoy}.
  \end{itemize}
    
  \item We also need to fix the initial condition $\pa B_{{\bf k}}^\pm/\pa t (t_i)$. To avoid an initial impulse of the magnetic helicity we set it to zero for all wave numbers. One can compute the value of the magnetic fields at $t_i+\Delta t$ as
  \be B_{{\bf k}}^\pm (t_i+\Delta t) \simeq B_{{\bf k}}^\pm (t_i) + \Delta t \cdot \frac{\pa B_{{\bf k}}^\pm}{\pa t} (t_i) \ee
  \item Then, one transforms Eq.~(\ref{AMW-1}) into a finite-difference equation using the approximation  
  \be  \frac{\pa^2 B_{{\bf k}}^\pm  }{\pa t^2} (t_i) \simeq \frac{B_{{\bf k}}^\pm (t_i+2\Delta t) - 2 B_{{\bf k}}^\pm (t_i+ \Delta t) + B_{{\bf k}}^\pm (t_i)   }{\Delta t^2} \ . \ee
   This allows us to obtain the magnetic field at $t_i+2\Delta t$ for all $k$
  \be B_{{\bf k}}^\pm (t_i+2 \Delta t) \simeq  \left( 2 -\sigma \Delta t \right) B_{{\bf k}}^\pm (t_i+\Delta t) - \left[ 1-\sigma \Delta t  +  \left(  k^2 \mp \frac{C \alpha \mu_5 (t_i) k}{\pi}\right) (\Delta t)^2  \right] B_{{\bf k}}^\pm (t_i) \ .    \ee
  This value is then used to extract the derivative of the fields at $t_i +\Delta t$.
  \be \frac{\pa B_{{\bf k}}^\pm}{\pa t} (t_i+\Delta t) \simeq  \frac{B_{{\bf k}}^\pm (t_i + 2 \Delta t) - B_{{\bf k}}^\pm (t_i)}{2 \Delta t} \ee
  \item At $t_i$ we compute the rhs of Eq.~(\ref{AMW-2}) by numerical integration of the magnetic fields and their derivatives. With this, we obtain $dn_5/dt (t_i)$ to predict
  \be n_5 (t_i+\Delta t) \simeq n_5 (t_i) + \Delta t \cdot \frac{d n_5}{dt} (t_i) \ee
  \item Using Eq.~(\ref{eq:n5mu5}) one can convert $n_5 (t_i+\Delta t)$ into $\mu_5(t_i+\Delta t)$. 
  \item We iterate the last three steps until the final time $t_f$.
 \end{enumerate}
 
  We need to probe several orders of magnitude in time. For this reason, we use a dynamical discretization step $\Delta t$ that increases with time to speed up the calculation.

\section{Helicity flip scattering\label{app:fliprate}}

In this Appendix we justify the reason why one can neglect the helicity-flipping rate $\Gamma_f$ that enters into Eq.~(\ref{anomal-full})
for the QGP in the time scales we are considering the system.

There are several scattering processes that allow us to change the quark helicity. Let us first discuss Compton scattering in detail,
when a massive fermion might change its helicity due to Compton scattering~\cite{Peskin:1995ev}. For almost massless fermions, helicity becomes synonymous with
chirality.  For simplicity, let us consider the situation where  $\mu_R,\mu_L \ll T$, and compute the rate in QED. We compute the helicity-flipping rate associated to Compton scattering as
\be \label{eq:gamma} \Gamma_f = - \frac{1}{n_5} \frac{dn_5}{dt} = - \frac{6}{T^2 \mu_5} \left( \frac{dn_R}{dt} - \frac{dn_L}{dt} \right) \ . \ee
neglecting completely the effect of the quantum anomaly. We compute the rhs of the Eq.~(\ref{eq:gamma}) with the use of the Boltzmann--Uehling--Uhlenbeck equation for right-handed fermions
\be
 \frac{df_{\rm FD}^R (t,p)}{dt}  = \int \frac{d^3k}{(2\pi)^3 2E_k} \frac{d^3 k'}{(2\pi)^3 2E_{k'}} \frac{d^3 p'}{(2\pi)^3 2E_{p'}}  \ \frac{1}{2E_p} (2\pi)^4 \delta^{(4)} (p+k-p'-k') |{\cal M}|^2  F \ ,
 \label{eq:gammaf}
 \ee
  where $F$ denotes the combination of distribution functions
\be 
F = f_{\rm FD} (p') f_{\rm BE} (k') [ 1+f_{\rm BE} (k)] [1-f_{\rm FD} (p)] - f_{\rm BE}(k) f_{\rm FD}(p) [1-f_{\rm FD}(p')][1+ f_{\rm BE}(k') ] 
\ee
and $f_{\rm FD}$ and $f_{\rm BE}$ stand for the Fermi--Dirac and Bose--Einstein thermal equilibrium distributions corresponding to the electrons and photons, respectively
We integrate over ${\bf p}$ to obtain the rate $dn_R/dt$,
\be \label{eq:dnrdt} \frac{dn_R}{dt}  = \int \frac{d^3p}{(2\pi)^3} \frac{df_R}{dt} \ . \ee
In the computation we ignore all the fermion masses in the kinematic factors, e.g. $E(p) \simeq p$ and concentrate on the leading term, in the
scattering amplitude squared, which reads~\cite{Peskin:1995ev}
\be |{\cal M}_{e_R^- \gamma_L \rightarrow e^-_L \gamma_R}|^2 = 4 e^4 \frac{m^2 s}{(-u +m^2)^2} \ , \ee
where the mass in the denominator is kept to regularize the infrared limit of the helicity flip rate, and the expression has been written in terms of the Mandelstam variables.

Given this expression we follow the ``$u$-channel parametrization'' method in Ref.~\cite{Arnold:2003zc} to simplify Eq.~(\ref{eq:dnrdt}).
The integration over ${\bf p'}$ is trivially performed by the 3-momentum conservation. The outgoing momentum
${\bf k'}$ is traded by ${\bf q}={\bf k'}-{\bf p}$, which is taken along the OZ axis, and a rigid rotation of $4\pi$ gives two more integrations. The
momentum ${\bf p}$ is taken to be in the OXZ plane with an angle $\theta_{pq}$ with respect to the $Z$ axis, and its
azimuthal integration gives a trivial $2\pi$. Finally, ${\bf k}$ is arbitrary in space, parametrized by its modulus
and its polar $\theta_{kq}$ and azimuthal angle $\phi$.

We further introduce the energy transfer variable $\omega$,
\be \label{eq:deltas} \delta (p+k-p'-k') = \int_{-\infty}^{\infty} \ d\omega \ \delta(\omega+p-k') \ \delta(\omega-k+p') \ . \ee
The two Dirac deltas in (\ref{eq:deltas}) help to perform the polar integrals using
\ba \delta (\omega +p-k') &=& \frac{k'}{pq} \delta \left( \cos \theta_{pq} -  \frac{u}{2pq} - \frac{ \omega }{q}\right) \Theta (\omega+p) \ , \\
 \delta (\omega -k+p') &=& \frac{p'}{kq} \delta \left( \cos \theta_{kq} +  \frac{u}{2kq} - \frac{ \omega }{q}\right) \Theta (k-\omega) \ . \ea
These changes bring the following conditions for the integration variables:
\be p > \frac{1}{2} (q-\omega) \ ; \ k > \frac{1}{2} (q+\omega) \ ; \ |\omega | < q \ . \ee

Introducing all terms, we can express~(\ref{eq:dnrdt}) as a five-dimensional integral:
\be \frac{dn_R}{dt} = \frac{e^4 m^2 }{2 (2\pi)^6} \int_0^{\infty} dq \int_{-q}^q d\omega \int^{\infty}_{\frac{q-\omega}{2}} dp \int_{\frac{q+\omega}{2}}^\infty  dk \int_0^{2\pi}  d\phi  
\ \frac{s}{(\omega^2 - q^2  -m^2)^2} \ F \ . \ee
The integral in $\phi$ can be easily performed by noting that only the $s$ variable depends on it $s=s(\cos \phi)$. The rate is now reduced to four integrals.
\be \frac{dn_R}{dt} =   \frac{e^4 m^2 \pi }{2 (2\pi)^6} \int_0^{\infty} dq \int_{-q}^q d\omega \int^{\infty}_{\frac{q-\omega}{2}} dp \int_{\frac{q+\omega}{2}}^\infty  dk \
\frac{\omega^2 -q^2 -4pk +2p\omega -2k\omega }{(\omega^2 - q^2  -m^2)^2} \ \frac{\omega^2-q^2}{q^2}  \ F \ . \ee

The Fermi--Dirac distribution functions are expanded to linear order in the chemical potential so that $F$ can be linearized in $\mu_5$,
\be F = - \frac{\mu_5}{T} \frac{1}{e^{\beta(k-\omega)}+1}  \frac{1}{e^{\beta (p+\omega)}-1} \frac{e^{\beta k}}{e^{\beta k}-1}
\frac{e^{\beta p}}{e^{\beta p}+1} + {\cal O}  \left(\frac{\mu_{R,L}^2}{T^2} \right) \ . \ee 

Finally, we express the solution in terms of the adimensional variables $q\beta = x_1 \ ; \ \omega\beta = x_2 \ ;
 \ p\beta=x_3 \ ; \ k\beta =x_4 \ , m \beta=y$ to obtain
\be \frac{dn_R}{dt} =  \frac{e^4 m^2 \mu_5 T }{128 \pi^5} \ \gamma(y^2) \ee
where we have defined the function 
\ba \gamma(y^2) & \equiv & \int_0^{\infty} dx_1 \int_{-x_1}^{x_1} dx_2 \int^{\infty}_{\frac{x_1-x_2}{2}} dx_3 \int_{\frac{x_1+x_2}{2}}^\infty  dx_4 \
 \nn \\
& \times& \frac{x_2^2 -x_1^2 -4x_3x_4 +2x_3x_2 -2x_4 x_2 }{(x_2^2 - x_1^2  -y^2)^2} \ \frac{x_2^2-x_1^2}{x_1^2}  
\frac{1}{e^{x_4-x_2}+1}  \frac{1}{e^{ x_3+x_2 }-1} \frac{e^{x_4}}{e^{x_4} -1} \frac{e^{x_3}}{e^{x_3}+1}  
   \nn \ . \ea

Adding a similar contribution for the left-handed fermion to get $dn_5/dt$ and using Eq.~(\ref{eq:gamma}), we obtain
\be  \Gamma_f =  \frac{3}{2\pi^3} \alpha^2 \frac{m^2}{T} \gamma(y^2) \ .  \ee
 The function $\gamma(y^2)$ is to be computed numerically and it acts as an IR regulator for the $\Gamma_f$, due to the fact that
it diverges in the limit $y \rightarrow 0$. This function is shown in Fig.~\ref{fig:gamma}.
\begin{figure}[h]
\begin{center}
\includegraphics[scale=0.4]{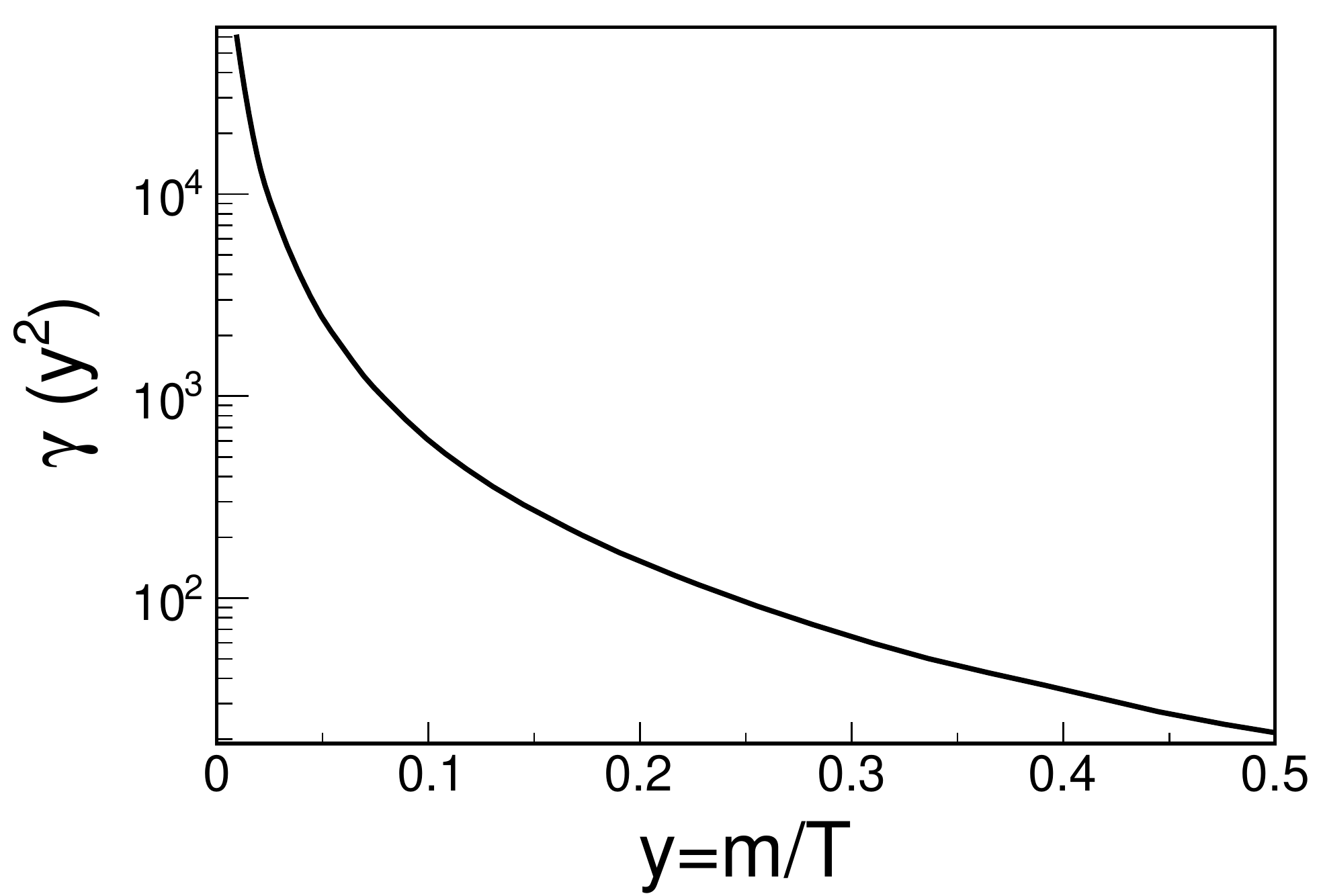}
\end{center} 
\caption{\label{fig:gamma} Function $\gamma(y^2)$ as a function of $y=m/T$.}
\end{figure}

Fixing $T=225$ MeV and $\alpha=1/137$ we find
\be
\tau^{QED}_f = \Gamma_f^{-1} \simeq \left\{ 
\begin{array}{rl} 
 5.6 \cdot 10^4 & \textrm{fm (for } m=5 \textrm{ MeV)} \ , \\
 6.4 \cdot 10^4 & \textrm{fm (for } m=100 \textrm{ MeV)} \ ,
 \end{array} \right.
 \ee
 while for $T= 500$ MeV we get 
\be
\tau^{QED}_f = \Gamma_f^{-1} \simeq \left\{ 
\begin{array}{rl} 
 2.5 \cdot 10^4 & \textrm{fm (for } m=5 \textrm{ MeV)} \ , \\
 2.6 \cdot 10^4 & \textrm{fm (for } m=100 \textrm{ MeV)} \ .
 \end{array} \right.
 \ee

Evidently, the quark scattering with polarized photons cannot give any contribution to the helicity flip rate in the typical time scales for HICs.
Results for the quark-gluon scattering) can be obtained from the QED result by changing $\alpha \rightarrow \alpha_s$ and multiplying the rate by the color
factor $2/9$ ~\cite{Peskin:1995ev}. For $T=225$ MeV
\be \tau^{QCD}_f = \tau^{QED}_f \ \frac{9 \alpha^2}{2 \alpha_s^2} \simeq  \left\{
\begin{array}{rl} 
  121 & \textrm{fm (for } m=5 \textrm{ MeV)} \ , \\
 138 & \textrm{fm (for } m=100 \textrm{ MeV)} \ , 
\end{array} \right. 
\ee
whereas for $T=500$ MeV,
\be \tau^{QCD}_f \simeq  \left\{
\begin{array}{rl} 
  54 & \textrm{fm (for } m=5 \textrm{ MeV)} \ , \\
 57 & \textrm{fm (for } m=100 \textrm{ MeV)} \ .
\end{array} \right. 
\ee
These times have been reduced by the strong coupling constant, but they are still out of the limits of the lifetime of a HIC fireball, and
clearly much longer than the time scales obtained for the change of the chirality associated to the presence chiral anomaly.

Quark-quark scattering events might also gives a contribution to $\Gamma_f$, but they should also scale as $\alpha^2_s m^2/T$, and although
we have not explicitly computed them, we expect them to be also irrelevant for the change of the chiral imbalance in the HIC.

Finally, we notice that the helicity change of a fermion (either massless or massive) by Landau damping is ruled out.
This process can be described by cutting the fermion self-energy diagram (Fig.~\ref{fig:landau}) at finite temperature and/or
density~\cite{Manuel:2000mk} (see also Ref.~\cite{LeBellac:1996kr}).
\begin{figure}[h]
\begin{center}
\includegraphics[scale=0.4]{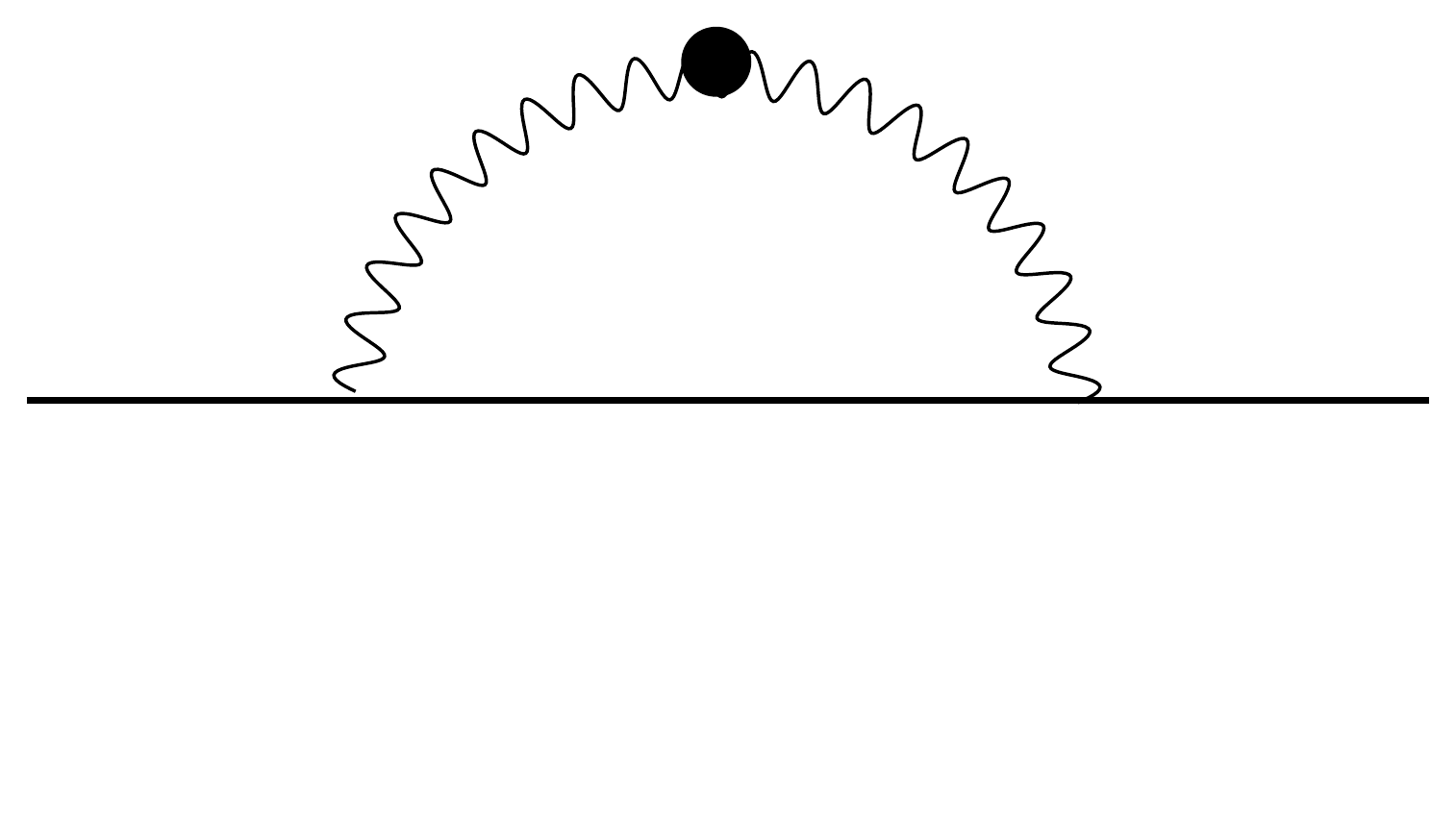}
\end{center} 
\caption{\label{fig:landau} Fermion self-energy diagram, the imaginary part of which encodes Landau damping.}
\end{figure}

The damping rate is defined in terms of the imaginary part of the fermion self-energy evaluated on shell ($E^2=p^2+m^2$)
\be \gamma(E)=\left. -\frac{1}{4E} \textrm{ Tr } [ \textrm{Im } \Sigma(p_0+i\eta,{\bf p}) (\slashed{P}+m)] \right|_{p_0=E} \ , \ee
where the fermion self-energy can be computed in the imaginary time formalism
\be \Sigma(i\omega_n,{\bf p}) = e^2 T \sum \int \frac{d^3q}{(2\pi)^3} \gamma_\mu S (P-Q) \gamma_\nu 
\Delta_{\mu \nu} (Q)
\ , \ee
with $S(P-Q)$ and $\Delta_{\mu \nu}(Q)$ the fermion free propagator and the photon polarization function, respectively.

Using the right-/left-handed helicity projectors defined in Ref.~\cite{Manuel:2000mk},
\be {\cal P}^{\pm} ({\bf p})= \frac{1\pm \gamma_5 \gamma_0 {\bf \gamma} \cdot \hat{\bf p}}{2} \ , \ee

one can compute the helicity-flipping damping rate as
\be \gamma(E)=\left. -\frac{1}{4E} \textrm{ Tr } [ \textrm{Im } \ {\cal P}^+ ({\bf p}) \
\Sigma(p_0+i\eta,{\bf p}) \ {\cal P}^- ({\bf p}) \ (\slashed{P}+m)] \right|_{p_0=E} \ . \ee

Using the algebra of Dirac matrices and the cyclic property of the trace one can easily check that the Dirac
trace vanishes for a helicity-changing process. Therefore, Landau damping cannot contribute to the 
helicity change of a fermion.


\begin{thebibliography}{99}


\bibitem{Kharzeev:2013ffa} 
  D.~E.~Kharzeev,
  Prog.\ Part.\ Nucl.\ Phys.\  {\bf 75}, 133 (2014)
  [arXiv:1312.3348 [hep-ph]].

\bibitem{Vilenkin:1980fu} 
  A.~Vilenkin,
  Phys.\ Rev.\ D {\bf 22}, 3080 (1980).
 
\bibitem{Kharzeev:2007jp} 
  D.~E.~Kharzeev, L.~D.~McLerran and H.~J.~Warringa,
  Nucl.\ Phys.\ A {\bf 803}, 227 (2008)
  [arXiv:0711.0950 [hep-ph]].



\bibitem{Fukushima:2008xe} 
  K.~Fukushima, D.~E.~Kharzeev and H.~J.~Warringa,
  Phys.\ Rev.\ D {\bf 78}, 074033 (2008)
  [arXiv:0808.3382 [hep-ph]].



\bibitem{Son:2012bg} 
  D.~T.~Son and B.~Z.~Spivak,
  Phys.\ Rev.\ B {\bf 88}, 104412 (2013)
  [arXiv:1206.1627 [cond-mat.mes-hall]].

\bibitem{Basar:2013iaa} 
  G.~Basar, D.~E.~Kharzeev and H.~U.~Yee,
  Phys.\ Rev.\ B {\bf 89}, no. 3, 035142 (2014)
  [arXiv:1305.6338 [hep-th]].

\bibitem{Li:2014bha} 
  Q.~Li, D.~E.~Kharzeev, C.~Zhang, Y.~Huang, I.~Pletikosic, A.~V.~Fedorov, R.~D.~Zhong and J.~A.~Schneeloch {\it et al.},
  arXiv:1412.6543 [cond-mat.str-el].

\bibitem{Gorbar:2011ya} 
  E.~V.~Gorbar, V.~A.~Miransky and I.~A.~Shovkovy,
  Phys.\ Rev.\ D {\bf 83}, 085003 (2011)
  [arXiv:1101.4954 [hep-ph]].


\bibitem{Yee:2009vw} 
  H.~U.~Yee,
  JHEP {\bf 0911}, 085 (2009)
  [arXiv:0908.4189 [hep-th]].

\bibitem{Rebhan:2009vc} 
  A.~Rebhan, A.~Schmitt and S.~A.~Stricker,
  JHEP {\bf 1001}, 026 (2010)
  [arXiv:0909.4782 [hep-th]].

\bibitem{Gynther:2010ed} 
  A.~Gynther, K.~Landsteiner, F.~Pena-Benitez and A.~Rebhan,
  JHEP {\bf 1102}, 110 (2011)
  [arXiv:1005.2587 [hep-th]].

\bibitem{Buividovich:2009wi} 
  P.~V.~Buividovich, M.~N.~Chernodub, E.~V.~Luschevskaya and M.~I.~Polikarpov,
  Phys.\ Rev.\ D {\bf 80}, 054503 (2009)
  [arXiv:0907.0494 [hep-lat]].

\bibitem{Abramczyk:2009gb} 
  M.~Abramczyk, T.~Blum, G.~Petropoulos and R.~Zhou,
  PoS LAT {\bf 2009}, 181 (2009)
  [arXiv:0911.1348 [hep-lat]].

\bibitem{Son:2012wh} 
  D.~T.~Son and N.~Yamamoto,
  Phys.\ Rev.\ Lett.\  {\bf 109}, 181602 (2012)
  [arXiv:1203.2697 [cond-mat.mes-hall]].
  
\bibitem{Stephanov:2012ki} 
  M.~A.~Stephanov and Y.~Yin,
  Phys.\ Rev.\ Lett.\  {\bf 109}, 162001 (2012)
  [arXiv:1207.0747 [hep-th]].
  


  
\bibitem{Son:2012zy} 
  D.~T.~Son and N.~Yamamoto,
  Phys.\ Rev.\ D {\bf 87}, no. 8, 085016 (2013)
  [arXiv:1210.8158 [hep-th]].

\bibitem{Manuel:2013zaa} 
  C.~Manuel and J.~M.~Torres-Rincon,
  Phys.\ Rev.\ D {\bf 89}, 096002 (2014)
  [arXiv:1312.1158 [hep-ph]].


  
\bibitem{Manuel:2014dza} 
  C.~Manuel and J.~M.~Torres-Rincon,
  Phys.\ Rev.\ D {\bf 90}, 076007 (2014)
  [arXiv:1404.6409 [hep-ph]].


\bibitem{Giovannini:1997eg} 
  M.~Giovannini and M.~E.~Shaposhnikov,
  Phys.\ Rev.\ D {\bf 57}, 2186 (1998)
  [hep-ph/9710234].
  
 
  
 

\bibitem{Boyarsky:2011uy} 
  A.~Boyarsky, J.~Frohlich and O.~Ruchayskiy,
  Phys.\ Rev.\ Lett.\  {\bf 108}, 031301 (2012)
  [arXiv:1109.3350 [astro-ph.CO]].

\bibitem{Tashiro:2012mf} 
  H.~Tashiro, T.~Vachaspati and A.~Vilenkin,
  Phys.\ Rev.\ D {\bf 86}, 105033 (2012)
  [arXiv:1206.5549 [astro-ph.CO]].


\bibitem{Banerjee:2004df} 
  R.~Banerjee and K.~Jedamzik,   
  Phys.\ Rev.\ D {\bf 70}, 123003 (2004)
  [astro-ph/0410032].


\bibitem{Jackson}
J.~D.~Jackson, ``Classical Electrodynamics" (John Wiley and Sons, New York, 1975).

\bibitem{Kharzeev:2010gr} 
  D.~E.~Kharzeev and D.~T.~Son,
  Phys.\ Rev.\ Lett.\  {\bf 106}, 062301 (2011)
  [arXiv:1010.0038 [hep-ph]].




\bibitem{Skokov:2009qp} 
  V.~Skokov, A.~Y.~Illarionov and V.~Toneev,
  Int.\ J.\ Mod.\ Phys.\ A {\bf 24}, 5925 (2009)
  [arXiv:0907.1396 [nucl-th]].



\bibitem{Peskin:1995ev} 
  M.~E.~Peskin and D.~V.~Schroeder,
  ``An Introduction to quantum field theory,''
  Reading, USA: Addison-Wesley (1995) 842 p.



\bibitem{Akamatsu:2013pjd} 
  Y.~Akamatsu and N.~Yamamoto,
  Phys.\ Rev.\ Lett.\  {\bf 111}, 052002 (2013)
  [arXiv:1302.2125 [nucl-th]].



\bibitem{Akamatsu:2014yza} 
  Y.~Akamatsu and N.~Yamamoto,
  Phys.\ Rev.\ D {\bf 90}, no. 12, 125031 (2014)
  [arXiv:1402.4174 [hep-th]].

\bibitem{Arnold:2005vb} 
  P.~B.~Arnold, G.~D.~Moore and L.~G.~Yaffe,
  Phys.\ Rev.\ D {\bf 72}, 054003 (2005)
  [hep-ph/0505212].
  


\bibitem{McLerran:1990de} 
  L.~D.~McLerran, E.~Mottola and M.~E.~Shaposhnikov,
  Phys.\ Rev.\ D {\bf 43}, 2027 (1991).



\bibitem{Moore:2010jd} 
  G.~D.~Moore and M.~Tassler,
  JHEP {\bf 1102}, 105 (2011)
  [arXiv:1011.1167 [hep-ph]].
  
\bibitem{Son:2002sd} 
  D.~T.~Son and A.~O.~Starinets,
  JHEP {\bf 0209}, 042 (2002)
  [hep-th/0205051].
  


\bibitem{Fukushima:2010vw} 
  K.~Fukushima, D.~E.~Kharzeev and H.~J.~Warringa,
  Phys.\ Rev.\ Lett.\  {\bf 104}, 212001 (2010)
  [arXiv:1002.2495 [hep-ph]].


\bibitem{Manuel:2004gk} 
  C.~Manuel and S.~Mrowczynski,
  Phys.\ Rev.\ D {\bf 70}, 094019 (2004)
  [hep-ph/0403024].




\bibitem{Kharzeev:2001ev} 
  D.~Kharzeev, A.~Krasnitz and R.~Venugopalan,
  Phys.\ Lett.\ B {\bf 545}, 298 (2002)
  [hep-ph/0109253].

\bibitem{Lappi:2006fp} 
  T.~Lappi and L.~McLerran,
  Nucl.\ Phys.\ A {\bf 772}, 200 (2006)
  [hep-ph/0602189].
  


\bibitem{Hirono:2014oda} 
  Y.~Hirono, T.~Hirano and D.~E.~Kharzeev,
  arXiv:1412.0311 [hep-ph].

  
\bibitem{Tuchin:2014iua} 
  K.~Tuchin,
  Phys.\ Rev.\ C {\bf 91}, no. 6, 064902 (2015)
  [arXiv:1411.1363 [hep-ph]].

  
\bibitem{Gursoy:2014aka} 
  U.~Gursoy, D.~Kharzeev and K.~Rajagopal,
  Phys.\ Rev.\ C {\bf 89}, 054905 (2014)
  [arXiv:1401.3805 [hep-ph]].

\bibitem{Ding:2010ga} 
  H.-T.~Ding, A.~Francis, O.~Kaczmarek, F.~Karsch, E.~Laermann and W.~Soeldner,
  Phys.\ Rev.\ D {\bf 83}, 034504 (2011)
  [arXiv:1012.4963 [hep-lat]].


\bibitem{Hongo:2013cqa} 
  M.~Hongo, Y.~Hirono and T.~Hirano,
  arXiv:1309.2823 [nucl-th].
  
\bibitem{Candelaresi:2011pg} 
  S.~Candelaresi and A.~Brandenburg,
  Phys.\ Rev.\ E {\bf 84}, 016406 (2011)
  [arXiv:1103.3518 [astro-ph.SR]].



\bibitem{Voronyuk:2011jd} 
  V.~Voronyuk, V.~D.~Toneev, W.~Cassing, E.~L.~Bratkovskaya, V.~P.~Konchakovski and S.~A.~Voloshin,
  Phys.\ Rev.\ C {\bf 83}, 054911 (2011)
  [arXiv:1103.4239 [nucl-th]].

\bibitem{Konchakovski:2012yg} 
  V.~P.~Konchakovski, E.~L.~Bratkovskaya, W.~Cassing, V.~D.~Toneev, S.~A.~Voloshin and V.~Voronyuk,
  Phys.\ Rev.\ C {\bf 85}, 044922 (2012)
  [arXiv:1201.3320 [nucl-th]].

\bibitem{Toneev:2011aa} 
  V.~D.~Toneev, V.~Voronyuk, E.~L.~Bratkovskaya, W.~Cassing, V.~P.~Konchakovski and S.~A.~Voloshin,
  Phys.\ Rev.\ C {\bf 85}, 034910 (2012)
  [arXiv:1112.2595 [hep-ph]].

\bibitem{Toneev:2012zx} 
  V.~D.~Toneev, V.~P.~Konchakovski, V.~Voronyuk, E.~L.~Bratkovskaya and W.~Cassing,
  Phys.\ Rev.\ C {\bf 86}, 064907 (2012)
  [arXiv:1208.2519 [nucl-th]].



\bibitem{Chen:2012ca} 
  J.~W.~Chen, S.~Pu, Q.~Wang and X.~N.~Wang,
  Phys.\ Rev.\ Lett.\  {\bf 110}, no. 26, 262301 (2013)
  [arXiv:1210.8312 [hep-th]].

\bibitem{Chen:2014cla} 
  J.~Y.~Chen, D.~T.~Son, M.~A.~Stephanov, H.~U.~Yee and Y.~Yin,
  Phys.\ Rev.\ Lett.\  {\bf 113}, no. 18, 182302 (2014)
  [arXiv:1404.5963 [hep-th]].
  
\bibitem{Bhatnagar:1954zz} 
  P.~L.~Bhatnagar, E.~P.~Gross and M.~Krook,
  Phys.\ Rev.\  {\bf 94}, 511 (1954).

\bibitem{Satow:2014lva} 
  D.~Satow and H.~-U.~Yee,
  Phys.\ Rev.\ D {\bf 90}, 014027 (2014)
  [arXiv:1406.1150 [hep-ph]].

\bibitem{Arnold:2003zc} 
  P.~B.~Arnold, G.~D.~Moore and L.~G.~Yaffe,
  JHEP {\bf 0305}, 051 (2003)
  [hep-ph/0302165].



\bibitem{Manuel:2000mk} 
  C.~Manuel,
  Phys.\ Rev.\ D {\bf 62}, 076009 (2000)
  [hep-ph/0005040].
  
\bibitem{LeBellac:1996kr} 
  M.~Le Bellac and C.~Manuel,
  Phys.\ Rev.\ D {\bf 55}, 3215 (1997)
  [hep-ph/9609369].

  
\end{thebibliography}
\end{document}